\documentclass[AMA,LATO1COL]{WileyNJD-v2}
\usepackage{moreverb}
\usepackage{placeins}
\usepackage{setspace}
\usepackage{amsmath}
\usepackage{amssymb}

\newcommand\BibTeX{{\rmfamily B\kern-.05em \textsc{i\kern-.025em b}\kern-.08em
T\kern-.1667em\lower.7ex\hbox{E}\kern-.125emX}}

\articletype{Research Article}%

\received{<day> <Month>, <year>}
\revised{<day> <Month>, <year>}
\accepted{<day> <Month>, <year>}

\begin{document}

\title{Data-driven modelling of turbine wake interactions and flow resistance in large wind farms}

\author[1]{Andrew Kirby*}

\author[2]{Fran\c{c}ois-Xavier Briol}

\author[3]{Thomas D. Dunstan}

\author[1]{Takafumi Nishino}

\authormark{KIRBY \textsc{et al}}

\address[1]{\orgdiv{Department of Engineering Science}, \orgname{University of Oxford}, \orgaddress{\state{Oxford}, \country{UK}}}

\address[2]{\orgdiv{Department of Statistical Science}, \orgname{University College London}, \orgaddress{\state{London}, \country{UK}}}

\address[3]{\orgdiv{Informatics Lab}, \orgname{UK MetOffice}, \orgaddress{\state{Exeter}, \country{UK}}}

\corres{*Andrew Kirby, Department of Engineering Science, University of Oxford, Oxford, OX1 3PJ, UK. \email{andrew.kirby@trinity.ox.ac.uk}}

\abstract[Abstract]{Turbine wake and local blockage effects are known to alter wind farm power production in two different ways: (1) by changing the wind speed locally in front of each turbine; and (2) by changing the overall flow resistance in the farm and thus the so-called farm blockage effect. To better predict these effects with low computational costs, we develop data-driven emulators of the `local' or `internal' turbine thrust coefficient $C_T^*$ as a function of turbine layout. We train the model using a multi-fidelity Gaussian Process (GP) regression with a combination of low (engineering wake model) and high-fidelity (Large-Eddy Simulations) simulations of farms with different layouts and wind directions. A large set of low-fidelity data speeds up the learning process and the high-fidelity data ensures a high accuracy. The trained multi-fidelity GP model is shown to give more accurate predictions of $C_T^*$ compared to a standard (single-fidelity) GP regression applied only to a limited set of high-fidelity data. We also use the multi-fidelity GP model of $C_T^*$ with the two-scale momentum theory (Nishino \& Dunstan 2020, J. Fluid Mech. 894, A2) to demonstrate that the model can be used to give fast and accurate predictions of large wind farm performance under various mesoscale atmospheric conditions. This new approach could be beneficial for improving annual energy production (AEP) calculations and farm optimisation in the future.}


\jnlcitation{\cname{%
\author{Kirby A.}, 
\author{Briol F-X.}, 
\author{Dunstan T.D.}, and 
\author{Nishino T.}} (\cyear{2022}), 
\ctitle{Data-driven modelling of wind turbine wake interactions in large wind farms}, \cjournal{Wind Energy}, \cvol{xxxx}.}

\maketitle

\section{Introduction}

\par The installed capacity of wind energy is projected to increase rapidly in the next decades. A major challenge in the optimisation of wind farm design is the accurate prediction of wind farm performance\cite{Porte-Agel2020}. Existing wind farm models struggle to make accurate predictions of wind farm power production. This is partly because the `global blockage effect' reduces the velocity upstream of large farms and hence the energy yield\cite{Bleeg2018}. It remains unclear how global blockage should be modelled and this is the subject of a large-scale field campaign\cite{CarbonTrust2022}. 

\par Wind farms are typically modelled using engineering `wake' models. These models predict the velocity deficit in the wakes behind turbines \cite{Jensen1983}\cite{Bastankhah2014}. To account for interactions between multiple turbines, the wake velocity deficits are superposed \cite{Katic1986,Zong2020}. Simple wake models can give predictions of wind farm performance with very low computational cost (~$10^{-3}$ CPU hours per simulation\cite{Porte-Agel2020}). However, wake models do not account for the response of the atmospheric boundary layer (ABL) to the wind farm which is likely to be important for large wind farms\cite{Kirby2022}. It has been found that wake models compare poorly to Large-Eddy Simulations (LES) of large wind farms \cite{Stevens2016a}.

\par Wind farms are also modelled in numerical weather prediction (NWP) models using farm parameterisation schemes. In these parameterisations, farms are often modelled as a momentum sink and a source of turbulent kinetic energy \cite{Fitch2012}. Turbine-wake interactions cannot be adequately predicted using these schemes. A new scheme was proposed\cite{Abkar2015} which uses a correction factor to model turbine interactions. More recently, data-driven approaches have been proposed\cite{Pan2018} to model these effects in wind farm parameterisations.

\par Data-driven modelling of wind farm flows is a promising new approach\cite{Zehtabiyan-Rezaie2022}. Data from high-fidelity simulations with complex flow physics can be used to make predictions with low computational cost. Recent studies have applied machine learning techniques to data from a single turbine or from an existing wind farm. The data for these studies are from measurements \cite{Renganathan2022,Optis2019,Japar2014,Yan2019}, LES \cite{Zhang2022} or Reynolds-Averaged Navier-Stokes (RANS) simulations \cite{Wilson2017,Ti2020,Ti2021}. A limitation of these approaches is that they are not generalisable to different turbine layouts unless they rely on wake superposition techniques to model farm flows. Another approach is modelling the effect of turbine layout using geometric parameters \cite{Yan2019} or using the layout as a graph input to a neural network \cite{Park2019, Bleeg2020}. However, these alternative approaches may struggle to fully capture the complex two-way interaction with the ABL as it seems impractical to prepare a data set that covers the entire range of scales involved in wind farm flows\cite{Porte-Agel2020}.

\par The problem of modelling wind farm flows can be split into `internal' turbine-scale and `external' farm-scale problems \cite{Nishino2020}. The `internal' problem is to determine a `local' or `internal' turbine thrust coefficient, $C_T^*$, which represents the flow resistance inside a wind farm, i.e., how the turbine thrust changes with wind speed within the farm. Nishino\cite{Nishino2016} proposed an analytical model for an upper limit of $C_T^*$ by using an analogy to the classic Betz analysis. This analytical model is a function of turbine-scale induction factor but is independent of turbine layout and wind direction. Previous studies \cite{Nishino2020} \cite{Nishino2016} \cite{Kirby2022} showed that $C_T^*$ is usually lower than the limit predicted by Nishino's model and can vary significantly with turbine layout due to wake and turbine blockage effects. 

\par The aim of this study is to develop statistical emulators of $C_T^*$ as a function of turbine layout and wind direction. The novelty of this approach is that we are modelling the effect of turbine-wake interactions on $C_T^*$ rather than turbine power. Both turbine-scale flows (e.g., wake effects) and farm-scale flows (e.g. farm blockage and mesoscale atmospheric response) affect turbine power within a farm. Therefore to create an emulator of turbine power, either (1) a very large set of expensive data such as finite-size wind farm LES is needed which covers a range of large-scale atmospheric conditions or (2) the model would not be generalisable to different mesoscale atmospheric responses. An emulator of $C_T^*$ is however applicable to different atmospheric responses modelled separately, following the concept of the two-scale momentum theory\cite{Nishino2020}\cite{Kirby2022}.

\par In section \ref{section:theory} we give the definitions of key wind farm parameters in the two-scale momentum theory \cite{Nishino2020}. Section \ref{section:simulations} summarises the methodology of the LES and wake model simulations, followed by the machine learning approaches to develop the emulators in section \ref{section:machine_learning}. In section \ref{section:results} we present the results from the trained emulators. These results are discussed in section \ref{section:discussion} and concluding remarks are given in section \ref{section:conclusion}.

\section{Two-scale momentum theory}\label{section:theory}

\par By considering the conservation of momentum for a control volume with and without a large wind farm over the land or sea surface, the following non-dimensional farm momentum (NDFM) equation can be derived\cite{Nishino2020},

\begin{equation}
    C_T^* \frac{\lambda}{C_{f0}} \beta^2 + \beta^\gamma = M
    \label{windfarmmomentum}
\end{equation}

\noindent where $\beta$ is the farm wind-speed reduction factor defined as $\beta\equiv U_F/U_{F0}$ (with $U_F$ defined as the average wind speed in the nominal wind farm-layer of height $H_F$, and $U_{F0}$ is the farm-layer-averaged speed without the wind farm present); $\lambda$ is the array density defined as $\lambda\equiv nA/S_F$ (where $n$ is the number of turbines in the farm, $A$ is the rotor swept area and $S_F$ is the farm footprint area); $C_T^*$ is the internal turbine thrust coefficient defined as $C_T^*\equiv\sum_{i=1}^{n}T_i/\frac{1}{2}\rho U_F^2nA$ (where $T_i$ is thrust of turbine $i$ in the farm and $\rho$ is the air density); $C_{f0}$ is the natural friction coefficient of the surface defined as $C_{f0}\equiv\langle\tau_{w0}\rangle/\frac{1}{2}\rho U_{F0}^2$ (where $\tau_{w0}$ is the bottom shear stress without the farm present); $\gamma$ is the bottom friction exponent defined as $\gamma\equiv\log_\beta (\langle \tau_w \rangle/ \tau_{w0})$ (where $\langle \tau_w \rangle$ is the bottom shear stress averaged across the farm); $M$ is the momentum availability factor defined as,

\begin{equation}
    M = \frac{\text{Momentum supplied by the atmosphere to the farm site }\textbf{with}\text{ turbines}}{\text{Momentum supplied by the atmosphere to the farm site }\textbf{without}\text{ turbines}}.
    \label{momentumavailability}
\end{equation}

\noindent noting that this includes pressure gradient forcing, Coriolis force, net injection of streamwise momentum through top and side boundaries and time-dependent changes in streamwise velocity\cite{Nishino2020}. The height of the farm-layer, $H_F$, is used to define the reference velocities $U_F$ and $U_{F0}$. Equation \ref{windfarmmomentum} is valid so long as the same of $H_F$ is used for both the internal and external problem. $H_F$ is typically between $2H_{hub}$ and $3H_{hub}$\cite{Kirby2022} (where $H_{hub}$ is the turbine hub-height) and in this study we use a fixed definition of $H_F=2.5H_{hub}$.  

\par Patel\cite{Patel2021} used an NWP model to demonstrate that, for most cases, $M$ varied almost linearly with $\beta$ (for a realistic range of $\beta$ between 0.8 and 1). Therefore, $M$ can be approximated by 

\begin{equation}
    M=1+\zeta (1-\beta)
    \label{extractability}
\end{equation}

\noindent where $\zeta$ is the `momentum response' factor or `wind extractability' factor. Patel\cite{Patel2021} found $\zeta$ to be time-dependent and vary between 5 and 25 for a typical offshore site (note that $\zeta=0$ corresponds to the case where momentum available to the farm site is assumed to be fixed, i.e., $M=1$).

\par Nishino\cite{Nishino2016} proposed an analytical model for $C_T^*$  given by,

\begin{equation}
    C_T^* = 4\alpha (1 - \alpha) = \frac{16C_T'}{(4+C_T')^2}
    \label{ctstar}
\end{equation}

\noindent where $\alpha$ is the turbine-scale wind speed reduction factor defined as $\alpha\equiv U_T/U_F$ ($U_T$ is the streamwise velocity averaged over the rotor swept area) and $C_T'\equiv T/\frac{1}{2}\rho U_T^2 A$ is a turbine resistance coefficient describing the turbine operating conditions.

\par For a given farm configuration at a farm site (i.e., for given set of $C_T^*$, $\lambda$, $C_{f0}$, $\gamma$ and $\zeta$) the farm wind-speed reduction factor $\beta$ can be calculated using equation \ref{windfarmmomentum}. The (farm-averaged) power coefficient $C_p$ is defined as $C_p\equiv \sum_{i=1}^{n}P_i/\frac{1}{2}\rho U_{F0}^3nA$ ($P_i$ is power of turbine $i$ in the farm). Using the calculated value of $\beta$, $C_p$ can be calculated by using the expression,  

\begin{equation}
    C_p = \beta^3 C_p^*
    \label{cp}
\end{equation}

\noindent where $C_p^*$ is the (farm-averaged) `local' or `internal' turbine power coefficient defined as $C_p^*\equiv \sum_{i=1}^{n}P_i/\frac{1}{2}\rho U_F^3nA$.

\section{Wind farm simulations}\label{section:simulations}

\FloatBarrier

\par
\begin{figure}
\centering
\includegraphics[width=0.8\textwidth]{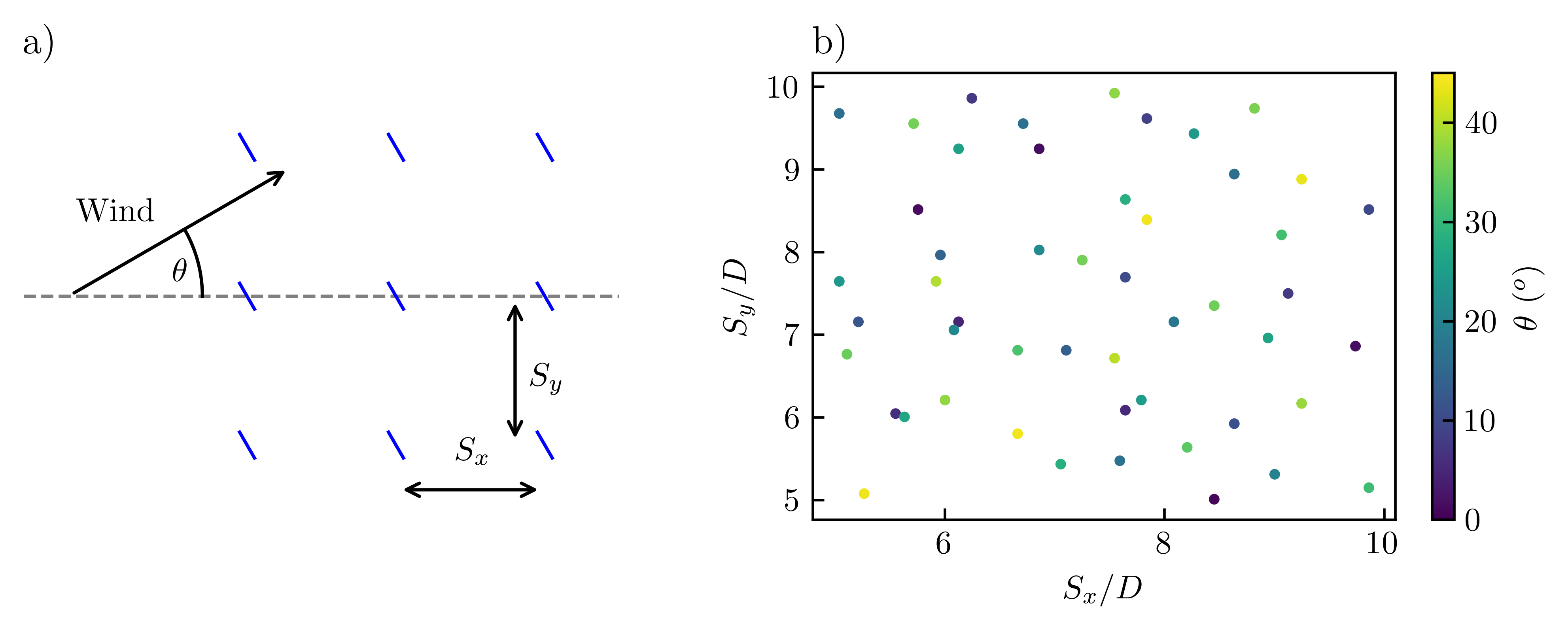}
\caption{Design of numerical experiments: a) input parameters, b) maximin design of LES.}
\label{fig:experimental_design}
\end{figure}

 In this study we model wind farms as arrays of actuator discs (or aerodynamically ideal turbines operating below the rated wind
speed). This is because, in real wind farms, the effects of turbine wake interactions on the farm performance are most significant when they operate below the rated wind speed. The `internal' thrust coefficient $C_T^*$ is an important wind farm parameter which includes the effect of turbine interactions (including both wake and local blockage effects). In this study we will be modelling the effect of turbine layout on $C_T^*$ for aligned turbine layouts with various wind directions and a fixed turbine resistance of $C_T'=1.33$. We chose $C_T'=1.33$ because it leads to a turbine induction factor of 1/4 which is close to a typical value for modern large wind turbines. As such we will be considering

\begin{equation}
       C_T^* = f(S_x, S_y, \theta)
        \label{ct*model}
\end{equation}

\noindent where $S_x$ is the turbine spacing in the $x$ direction, $S_y$ is the turbine spacing in the $y$ direction and $\theta$ is the wind direction relative to the $x$ direction (see figure \ref{fig:experimental_design}a). However the true function $C_T^*$ cannot be easily evaluated so we will instead investigate $C_T^*$ using computer codes. One computer code we will use is LES (see section \ref{les}) to estimate $C_T^*$

\begin{equation}
       C_{T,LES}^* = f_{LES}(S_x, S_y, \theta).
        \label{ct*les}
\end{equation}

\noindent We assume that the function $f_{LES}$ is close to the true function $f$ because of the accuracy of LES to model wind farm flows. We will also use a wake model (see section \ref{wake_model}) to provide cheap approximations of $C_T^*$ according to 

\begin{equation}
       C_{T,wake}^* = f_{wake}(S_x, S_y, \theta).
        \label{ct*wake}
\end{equation}

\par Engineering problems are often investigated using complex computer models. Evaluating the output of such computer models for a given input can be very computationally
expensive. Therefore a common objective is to create a cheap statistical model of the expensive computer model; this is commonly known as emulation of computer models \cite{Sacks1989}\cite{Currin1991}. In this study we aim to develop a statistical emulator which can cheaply emulate $f_{LES}$.

\par The emulators will only be valid for aligned layouts of wind turbines and for a given turbine resistance (here we use $C_T'=1.33$). We consider the input parameters for a realistic range of turbine spacings\cite{Porte-Agel2020}: $S_x \in [5 D, 10 D]$, $S_y \in [5 D, 10 D]$ and $\theta \in [0^o, 45^o]$ where $D$ is the diameter of the turbine rotor swept area. In this study $D$ is set as 100m and the turbine hub height is also 100m. We only need to consider wind directions of $\theta \in [0^o, 45^o]$ because of symmetry in the aligned turbine layouts. If $\theta$ is negative than the turbine layout given by $(S_x,S_y,\theta)$ is exactly the same as $(S_x,S_y,-\theta)$. When $\theta>45^o$, then $(S_x,S_y,\theta)$ and $(S_y,S_x,90^o-\theta)$ give identical layouts.

\par In this study we build several emulators to predict $f_{LES}$. The models are trained using data from low-fidelity (wake model) and high fidelity (LES) wind farm simulations. One evaluation of $C_{T,wake}^*$ takes approximately 130 seconds on a single CPU and $C_{T,LES}^*$ requires around 400 CPU hours on a supercomputer. We use a space filling maximin design \cite{Johnson1990}\cite{Santner2018} to select training points in the parameter space. The maximin algorithm selects points which maximises the minimum distance to other points and to the boundaries. This provides a good coverage of the domain which ensures that the emulators can give good predictions across the whole of the domain\cite{Wynne2021}. Figure \ref{fig:experimental_design}b shows the LES training points in the parameter space.

\FloatBarrier

\subsection{Large-Eddy Simulations}\label{les}

\begin{figure}
\centering
\includegraphics[width=0.8\textwidth]{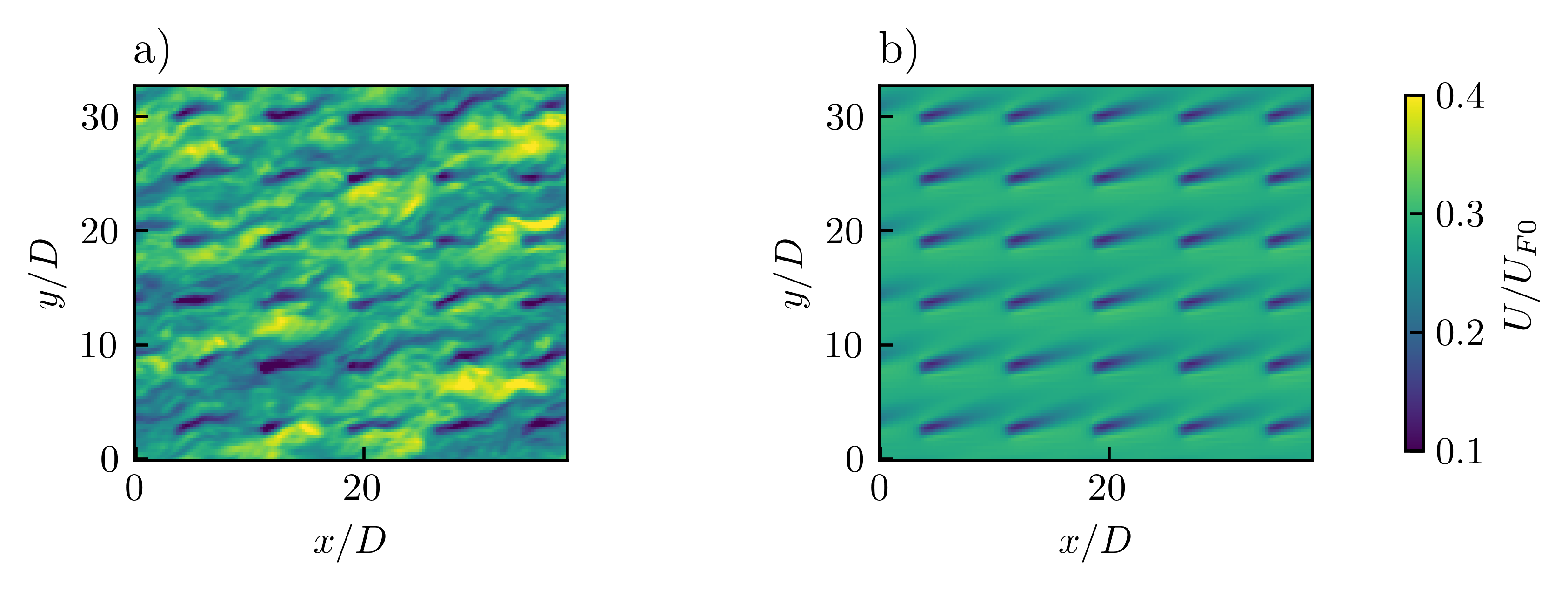}
\caption{LES a) instantaneous and b) time-averaged flow fields over a periodic turbine array ($S_x/D=7.59$, $S_y/D=5.47$ and $\theta=37.6^o$).}
\label{fig:les_results}
\end{figure}

This study uses the data from 50 high-fidelity (LES) simulations of wind farms published in a previous study\cite{Kirby2022}. Here we give a brief summary of the LES methodology. The LES models a neutrally stratified atmospheric boundary layer over a periodic array of actuator discs, which face the wind direction $\theta$ and exert uniform thrust. The resolution is 24.5m in the horizontal directions (4 points across the rotor diameter) and 7.87m in the vertical. This is a coarse horizontal resolution; however using a correction factor for the turbine thrust\cite{Shapiro2019} makes the $C_{T,LES}^*$ values insensitive to horizontal resolution\cite{Kirby2022}. For all simulations the vertical domain size was fixed at 1km and the horizontal extent varied with turbine layout but was at least 3.14km. The horizontal boundary conditions were periodic (essentially an infinitely-large wind farm). The bottom boundary used a no-slip condition with the value of eddy viscosity specified following the Monin-Obukhov similarity theory for a surface roughness length of $z_0=1\times10^{-4}$m. The top boundary had a slip condition with zero vertical velocity. The flow was driven by a pressure gradient forcing which was constant and in the direction $\theta$ throughout the domain. Figure \ref{fig:les_results} shows the instantaneous and time-averaged hub height velocities from one wind farm LES. See the original paper\cite{Kirby2022} for further details of the LES.

\FloatBarrier

\subsection{Wake model simulations}\label{wake_model}

\begin{figure}
\centering
\includegraphics[width=0.5\textwidth]{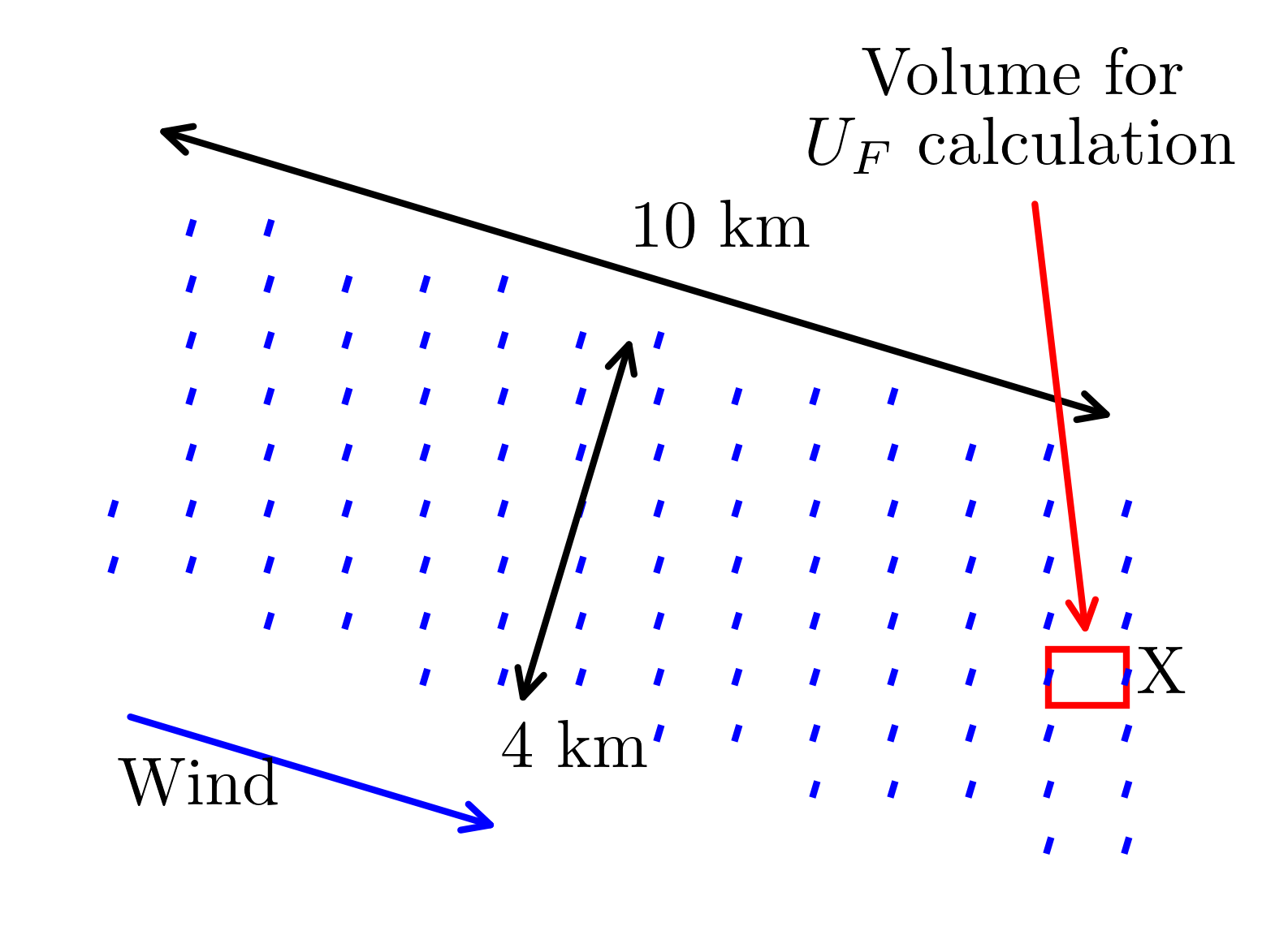}
\caption{Example of wind farm layout for wake model simulations.}
\label{fig:wake_model_setup}
\end{figure}

Wake models are a cheap low-fidelity approach to modelling wind farm aerodynamics compared to expensive high-fidelity LES simulations \cite{Porte-Agel2020}. We use the wake model proposed by Niayafar and Port\'e-Agel \cite{Niayifar2016} to evaluate $C_{T,wake}^*$ as a cheap approximation of $C_T^*$. We use the Python package PyWake \cite{pywake2.2.0_2020} to implement the wake model. The turbine thrust coefficient $C_T$ is needed as an input for the wake model. We use the value of $C_T^*$ predicted by equation \ref{ctstar} as the value of $C_T$. For the turbine operating conditions used in this study ($C_T'=1.33$) the wake model has $C_T$ equal to 0.75 for all turbines. To model actuator discs, we consider a hypothetical turbine which has a constant $C_T$ for all wind speeds. We calculate $C_{T,wake}^*$ for a single turbine at the back of a large farm (marked $X$ in figure \ref{fig:wake_model_setup}). The farm simulated using the wake model is 10km long in the streamwise direction and 4km long in the cross-streamwise direction. The farm size was chosen so that $C_T^*$ no longer varied with increasing farm size. The wake growth parameter is calculated using $k^*=0.38I+0.004$ where $I$ is the local streamwise turbulence velocity. The local streamwise turbulence intensity is estimated using the model proposed by Crespo and Hern\'andez\cite{Crespo1996}. The background turbulence intensity (TI) is set as a typical value of 10\%.

\par The velocity incident to the turbine is calculated by averaging the velocity across the disc area. We use a 4$\times$3 cartesian grid with Gaussian quadrature coordinates and weights on the disc to average the velocity. The disc-averaged velocity, $U_T$ is then calculated by multiplying the averaged incident velocity by $(1-a)$ where $a$ is the turbine induction factor set by the value of $C_T'$ (using the expression $a=C_T'/(4+C_T')$). To calculate the farm-average velocity, $U_F$, we average the velocity across a volume around the single turbine. The volume has dimensions of $S_y$ in the $y$ direction, $S_x$ in the $x$ direction and 250m in the $z$ direction (the height of the nominal farm layer used in the previous LES study\cite{Kirby2022}). To calculate the average velocity, we discretise the volume into 200 points in the horizontal directions and 20 points in the vertical. This was sufficient for the calculation of $C_{T,wake}^*$ to not vary with further discretisation. Figure \ref{fig:wake_model_setup} shows an example of the farm layout for the wake model simulations.

\section{Machine learning methodology}\label{section:machine_learning}

\subsection{Gaussian Process regression}\label{basic_gp}

\par We will use Gaussian process (GP) regression \cite{Rasmussen2018} to build statistical emulators of $f_{LES}$. A Gaussian process is a stochastic process $g\sim\mathcal{GP}(m,k)$ described by a mean function $m(v) = \mathbb{E}[g(v)]$ and a covariance function $k(v,v')=\mathbb{E}[(g(v)-m(v))(g(v')-m(v')]$. In our case $v=(S_x,S_y,\theta)$. We will use such a stochastic process as a model of $f_{LES}$, the true mapping from $v$ to $C_{T,LES}^*$. Each realisation from this process will therefore be a function which could plausibly represent this mapping. The mean function represents the expected output value at an input $v=(S_x,S_y,\theta)$. The covariance function gives the covariance between output values at $v$ and $v'$. Examples of covariance functions include squared exponential, rational quadratic and periodic functions\cite{Rasmussen2018}. Different covariance functions will give differently shaped GPs. For example the squared exponential covariance function will give very smooth GPs whereas the periodic function will give GPs with a periodic structure. Other types of structure, for example symmetry, can also be encoded in the covariance function. Therefore the expected shape (for example smoothness) of the expected relationship and any properties (for example discontinuities or symmetries) need to be considered when choosing a covariance function for GP regression.

\par Let $V = (v_1,...,v_n)^T$ be a collection of design points then $m_V = (m(v_1),...,m(v_n))^T$ is the mean vector and $k_{VV}=(k(v_i,v_j))$ is the covariance matrix. We will start by positing a GP model with mean $m$ and covariance $k$ (called the `prior GP'), then condition this GP on LES observations; the outcome is a new GP (called the `posterior GP'). This gives the posterior distribution $g|V,C_{T,LES}^* \sim \mathcal{GP}(\overline{m}_{\sigma^2},\overline{k}_{\sigma^2})$. $\overline{m}_{\sigma^2}$ is the posterior mean function given by $\overline{m}_{\sigma^2}(v)=m(v)+k_{vV}(k_{VV}+\sigma^2I_{n\times n})^{-1}(C_{T,LES}^*-m_V$) where $k_{vV}=(k(v,v_1),...,k(v,v_n))$ and $I_{n \times n}$ is the identity matrix of size $n$. The posterior mean function $\overline{m}_{\sigma^2}$ is used to make predictions at $v=(S_x, S_y, \theta)$. The posterior covariance function $\overline{k}_{\sigma^2}$ quantifies the uncertainty in our prediction at $v=(S_x, S_y, \theta)$. The posterior covariance function is given by $\overline{k}_{\sigma^2}(v,v')=k(v,v')-k_{vV}(k_{VV}+\sigma^2I_{n\times n})^{-1}k_{Vv'}$.

\par Often in GP regression a zero prior mean is used. However, using an informative prior mean can improve the accuracy of the trained model. By using a prior mean, many of the trends in $f_{LES}$ can be incorporated into our model prior to making expensive evaluations of $C_{T,LES}^*$. Therefore, after training our model will likely better describe the true relationship between $S_x,S_y,\theta$ and $f_{LES}$. In this study, we will use both $C_{T,wake}^*$ and the analytical model of $C_T^*$ as the prior mean for the standard GP regression. For the wake model prior mean we also vary the specified ambient TI input parameter.

\par We expect $f_{LES}$ to be a smooth function of input variables $S_x$, $S_y$ and $\theta$, and to vary more rapidly with $\theta$ than $S_x$ or $S_y$. Therefore we will use an anisotropic squared-exponential covariance function,

\begin{equation}
      k(v,v') = \sigma_f^2 \exp \left( -\frac{(S_x-S_x')^2}{2l_1^2}\right)\exp \left(-\frac{(S_y-S_y')^2}{2l_2^2}\right)\exp \left(-\frac{(\theta-\theta')^2}{2l_3^2}\right)
    \label{sekernel}
\end{equation}

\noindent where $\sigma_f^2>0$ is the signal variance hyperparameter and $l_i>0$ is the lengthscale hyperparameter for each dimension. This is also called an ARD (automatic relevance detection) kernel. If we consider $v=v'$ then we can see that $\sigma_f^2$ determines the variance of $g(v)$. Therefore $\sigma_f^2$ determines the prior uncertainty the model has about the value of $g(v)$. As the lengthscale hyperparameter $l_i$ gets smaller then $k(v,v')$ decreases (for $v\neq v'$). Equally if $l_i$ increases then $k(v,v')$ will also increase. A GP with a small $l_i$ will therefore vary more rapidly across the parameter space in the $i$th dimension.

\par Due to numerical issues associated with the matrix inversion/linear system solve operations in the formulae for the posterior GP, it is common to add a nugget $\sigma^2>0$ to the kernel matrix.  The hyperparameters $\sigma_f^2$ and $l_i$ are selected automatically during the fitting process by maximising the log marginal likelihood \cite{Rasmussen2018}. This approach selects the model which maximises the fit to the data.

\begin{figure}
\centering
\includegraphics[width=0.8\textwidth]{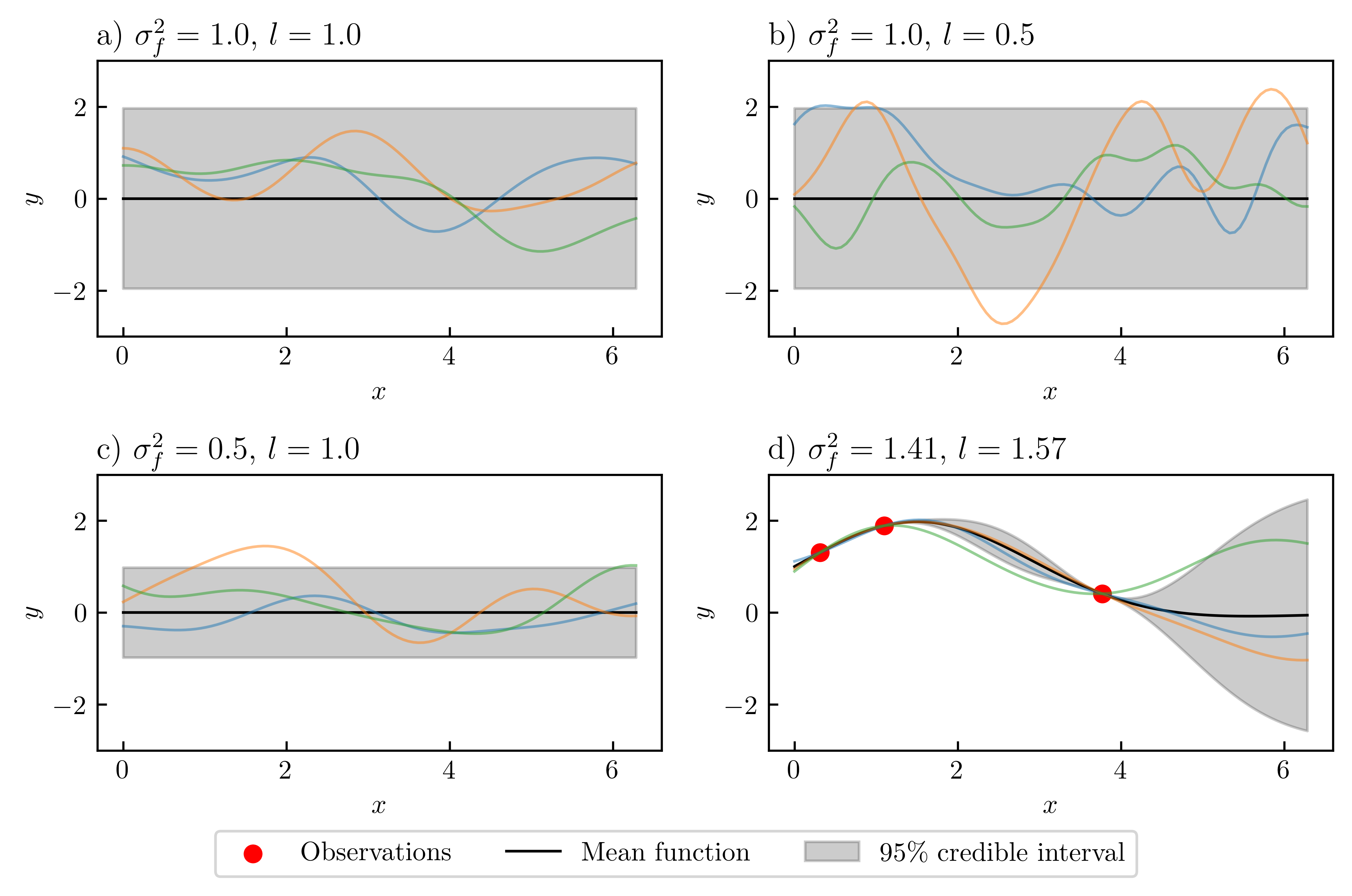}
\caption{Demonstration of basic GP regression: a) shows the prior mean and covariance function prior to fitting with 3 GPs drawn from the distribution shown in colour; b) shows the effect of decreasing the lengthscale hyperparameter; c) the effect of variance hyperparameter; and d) the posterior mean and covariance functions.}
\label{fig:gp_demo}
\end{figure}

\par Figure \ref{fig:gp_demo} shows the impact of the hyperparameters in an example GP regression setting (using the squared exponential covariance function). The mean function and 95\% credible interval (+/-1.96 times the standard deviation) prior to fitting are shown in figure \ref{fig:gp_demo}a with 3 GPs drawn from the distribution (coloured lines). The effect of decreasing the lengthscale hyperparameter $l_i$ is shown in figure \ref{fig:gp_demo}b. The prior mean and 95\% credible interval are unchanged however the example GPs drawn vary more rapidly because of the shorter lengthscale. Figure \ref{fig:gp_demo}c shows the same setup as figure \ref{fig:gp_demo}a but with a smaller value of $\sigma_f^2$. The example GPs still vary slowly but the magnitude of the variations is now smaller. Figure \ref{fig:gp_demo}d shows the GPs conditioned on observations with hyperparameters selected by maximising the log marginal likelihood.

\subsection{Non-linear multi-fidelity Gaussian Process regression}\label{mf_gp}

\par In many applications there are several computational models available. These models can have varying accuracies and computational costs. The models which are more computationally expensive typically give more accurate predictions. The GP regression framework can be extended to combine information from low and high-fidelity models \cite{Peherstorfer2018}. This type of modelling uses the low-fidelity observations to speed up the learning process and the high-fidelity observations to ensure accuracy. In our scenario we will combine evaluations of from a low-fidelity ($C_{T,wake}^*$) and a high-fidelity ($C_{T,LES}^*$) model. Note that for the multi-fidelity models in this study we set the ambient TI to 10\% for the wake model and use a zero prior mean. We will keep the number of high-fidelity training points fixed at 50 and we will vary the number of low-fidelity training points used.

\par We combine information from our high and low-fidelity models using a nonlinear information fusion algorithm \cite{Perdikaris2017}. The framework is based on the autoregressive multi-fidelity scheme given by:

\begin{equation}
      g_{high}(v) = \rho(g_{low}(v)) + \delta(v)
      \label{nonlinear_mf_GP}
\end{equation}

\noindent where $g_{low}(v)$ is a model with a GP denoted $f_{wake}$ and $g_{high}(v)$ is a model with a GP denoted $f_{LES}$. $\rho$ is a model with a GP which maps the low-fidelity output to the high-fidelity output and $\delta(v)$ is a model with a GP which is a bias term. The non-linear multi-fidelity framework can learn non-linear space-dependent correlations between models of different accuracies. To reduce the computational cost and complexity of implementation the autoregressive scheme given by equation \ref{nonlinear_mf_GP} is simplified. Firstly, the GP prior $g_{low}(v)$ is replaced by the GP posterior $g_{low,*}(v)$ and secondly the GPs $\rho$ and $\delta$ are assumed to be independent. Equation \ref{nonlinear_mf_GP} can then be summarised as

\begin{equation}
      g_{high}(v) = h_{high}(v,g_{low,*}(v))
      \label{nonlinear_mf_GP_simple}
\end{equation}

\noindent where $h_{high}$ is a model with a GP which has both $v$ and $g_{low,*}(v)$ as inputs. More details of $h_{high}$ and the implementation of the multi-fidelity framework are given in Perdikaris \textit{et. al.}\cite{Perdikaris2017}.

\begin{figure}
\centering
\includegraphics[width=0.8\textwidth]{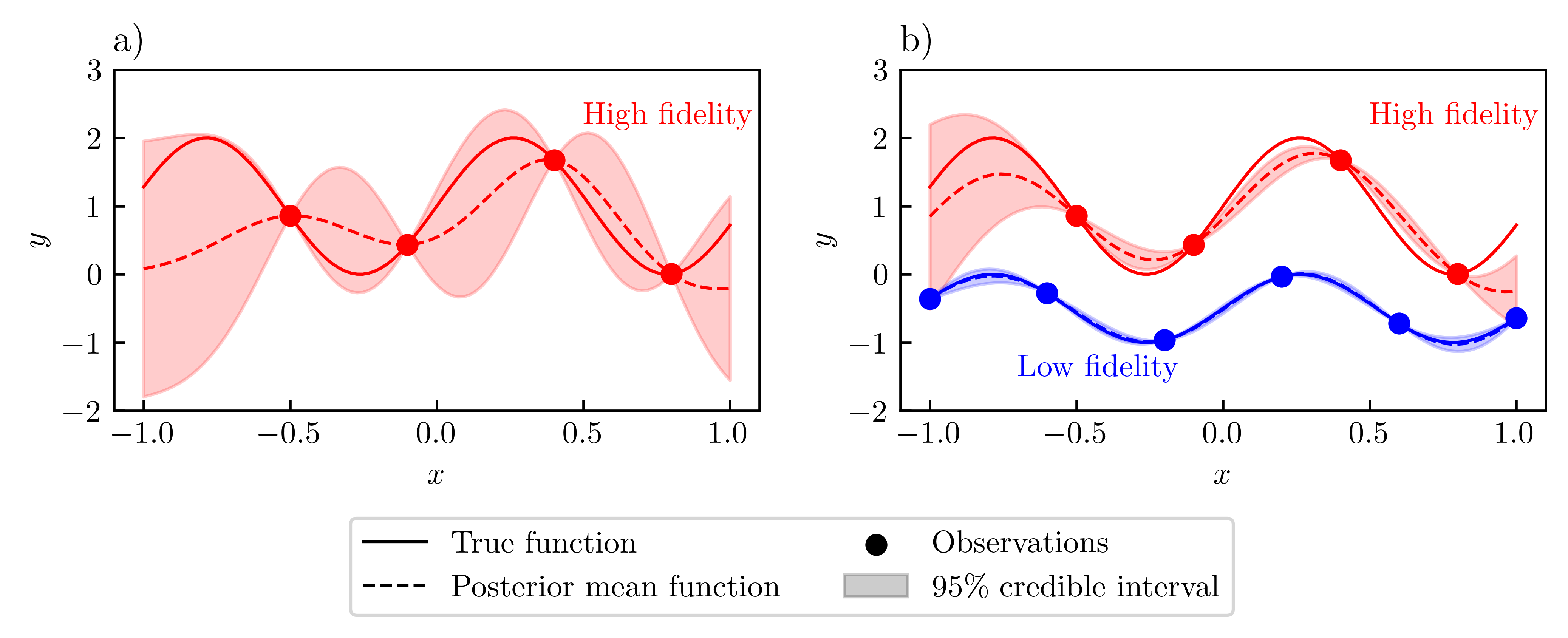}
\caption{Demonstration of a) basic GP regression and b) multi-fidelity GP regression. In this example $f(x)=1+sin(6x)$ for the high-fidelity data and $f(x)=-0.5+0.5sin(6x)$ for the low-fidelity data.}
\label{fig:mf_gp}
\end{figure}

\par Figure \ref{fig:mf_gp} shows an example of how a multi-fidelity GP can outperform a standard GP regression. We implement the non-linear multi-fidelity framework using the `emukit' package\cite{emukit2019}. We first maximise the log marginal likelihood whilst keeping the Gaussian noise variance fixed at a low value of $1\times10^{-6}$. The fitting process is then repeated whilst allowing the Gaussian noise variance to be optimised too. This is to prevent a high noise local optima from being selected.

\FloatBarrier

\section{Results}\label{section:results}

\par In this study, we build various statistical emulators of $f_{LES}$ using different techniques and compare the performance. A summary of the techniques is shown in the list below:

\begin{enumerate}[1]
    \item Standard Gaussian Process regression (see section \ref{basic_gp})
    \begin{enumerate}[a]
    \item \textbf{GP-analytical-prior}: Gaussian Process using analytical model (equation \ref{ctstar}) prior mean 
    \item \textbf{GP-wake-TI10-prior}: Gaussian Process using wake model (section \ref{wake_model}) with ambient TI=10\% prior mean
    \item \textbf{GP-wake-TI1-prior}: Gaussian Process using wake model with ambient TI=1\% prior mean
    \item \textbf{GP-wake-TI5-prior}: Gaussian Process using wake model with ambient TI=5\% prior mean
    \item \textbf{GP-wake-TI15-prior}: Gaussian Process using wake model with ambient TI=15\% prior mean
    \end{enumerate}
    \item Non-linear multi-fidelity Gaussian Process regression (see section \ref{mf_gp})
    \begin{enumerate}[a]
    \item \textbf{MF-GP-nlow500}: multi-fidelity Gaussian Process using 500 low-fidelity training points
    \item \textbf{MF-GP-nlow250}: multi-fidelity Gaussian Process using 250 low-fidelity training points
    \item \textbf{MF-GP-nlow1000}: multi-fidelity Gaussian Process using 1000 low-fidelity training points
    \end{enumerate}
\end{enumerate}

\par The code used to produce the results in this section is available open-access at the following GitHub repository: \url{https://github.com/AndrewKirby2/ctstar_statistical_model}.

\subsection{Performance of standard GP regression}\label{standardgp}

\FloatBarrier

\par We first assessed the accuracy of the standard GP models (section \ref{basic_gp}) by performing leave-one-out cross-validation (LOOCV). This is a method of estimating the accuracy of a statistical model when making predictions on data not used to train the model. We trained our model on 49 of the 50 training points and then calculated the prediction accuracy for the single high-fidelity data point which is excluded from the training set. This is then repeated for all data points in turn, and we took the average accuracy as an estimate of the model test accuracy. The standard GP models were implemented using the `GPy' package\cite{gpy2014}.

\par The standard GP gave accurate predictions of $f_{LES}$ with average errors of less than 2\%. Table \ref{tab:prior_means} shows the accuracy of the standard GP models compared to the analytical and wake models. We calculated the errors by using the expression $|\overline{m}_{\sigma^2}-C_{T,LES}^*|/0.75$ where $\overline{m}_{\sigma^2}$ is the posterior mean function of the emulator. The reference value for $C_T^*$ of 0.75 was chosen because this is the prediction from the analytical model. Both GP models give similar maximum errors of approximately 6\%. Using the wake model as a prior mean gave a lower mean absolute error of 1.26\%. The GP models reduced the average prediction error and significantly reduced the maximum error compared to the wake model and analytical model of $C_T^*$.

\begin{table}
\caption{Accuracy of models for $C_T^*$ prediction.}
\label{tab:prior_means}
\centering
\begin{tabular}{|c|c|c|}
\toprule
Model & MAE (\%) & Maximum error (\%)\\
\midrule
\textbf{GP-analytical-prior} & 1.87 & 6.09\\
\textbf{GP-wake-TI10-prior} & 1.26 & 6.11\\
Analytical model & 5.26 & 22.0\\
Wake model (TI=10\%) & 4.60 & 9.28\\
\bottomrule
\end{tabular}
\end{table}

\par
\begin{figure}
\centering
\includegraphics[width=0.8\textwidth]{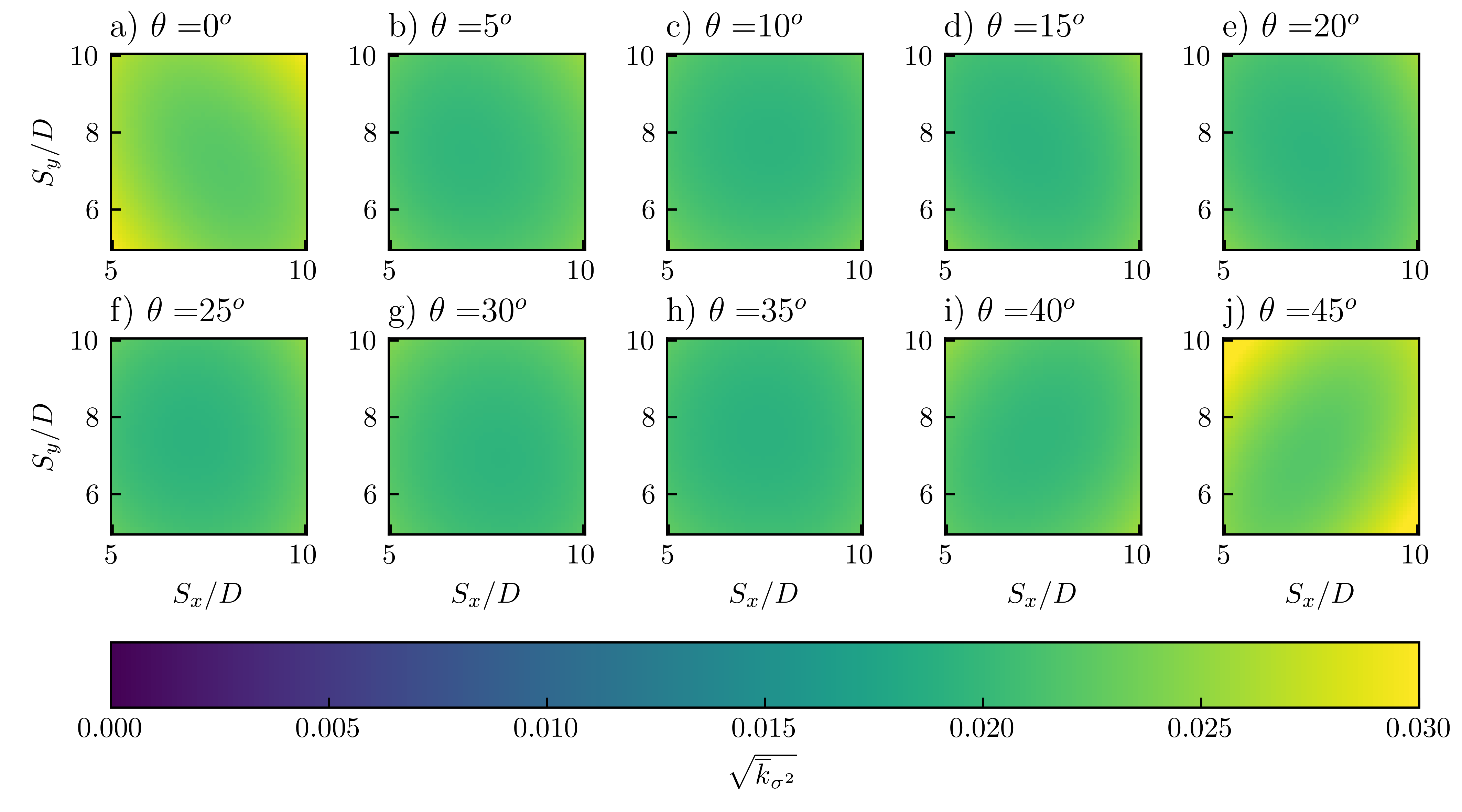}
\caption{Posterior variance function of $\textbf{GP-wake-TI10-prior}$ model.}
\label{fig:basic_gp_std}
\end{figure}

 The model $\textbf{GP-wake-TI10-prior}$ has a high degree of confidence when making predictions in regions of the parameter space. Figure \ref{fig:basic_gp_std} shows the square root of the posterior covariance function $\overline{k}_{\sigma^2}$, which quantities the uncertainty of the emulator. The uncertainty is uniform throughout the parameter space with regions of slightly higher uncertainty at $\theta=0^o$ and $45^o$. 

\par
\begin{figure}
\centering
\includegraphics[width=0.8\textwidth]{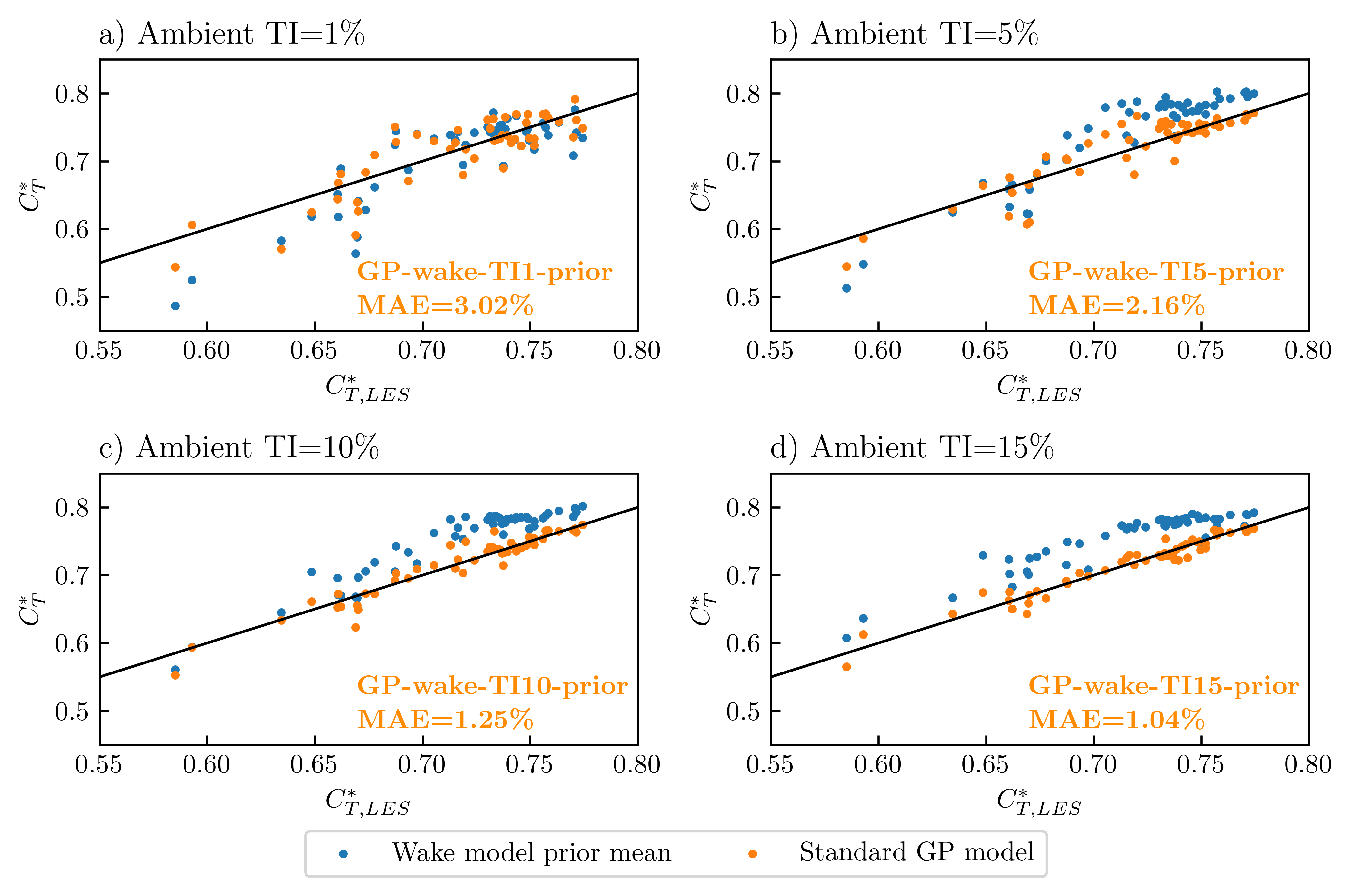}
\caption{Sensitivity of fitted GP models to the ambient TI chosen for wake model prior means.}
\label{fig:ti_sensitivity}
\end{figure}

 We also assessed the sensitivity of the model accuracy to the ambient TI used in the wake model prior mean. Figure \ref{fig:ti_sensitivity} shows the impact of ambient TI on the wake model prior mean and the fitted GP model. Increasing the ambient TI increased the value of $C_{T,wake}^*$. This is because of the enhanced wake recovery behind wind turbines. Increasing the ambient TI in the wake model results in $C_{T,wake}^*$ overpredicting $C_{T,LES}^*$. The MAE from the LOOCV procedure for each fitted GP is shown in the bottom right corner.

\par The fitted GPs became more accurate when the wake model ambient TI was increased. Increasing the ambient TI for the wake model causes the wakes to recover faster. The wakes become shorter in the streamwise direction and wider in the spanwise direction. As such, $C_{T,wake}^*$ becomes less sensitive to the turbine layout. When an ambient TI of 1\% and 5\% is used for the wake model, $C_{T,wake}^*$ is more sensitive to turbine layout than $C_{T,LES}^*$ (figures \ref{fig:ti_sensitivity}a and \ref{fig:ti_sensitivity}b). When the ambient TI is increased to 10\% and above, the relationship between $C_{T,wake}^*$ and $C_{T,LES}^*$ becomes simpler (figures \ref{fig:ti_sensitivity}c and \ref{fig:ti_sensitivity}d). This seems to explain why the fitted GPs become more accurate.

\FloatBarrier

\subsection{Performance of non-linear multi-fidelity GP regression}\label{results_ct*}

\begin{table}
\caption{Performance of the multi-fidelity Gaussian Process models.}
\label{tab:nlow_sensitivity}
\centering
\begin{tabular}{|c|c|c|c|c|}
\toprule
Model & MAE (\%) & Maximum error (\%) & Training time (s) & Prediction time (s)\\
\midrule
\textbf{MF-GP-nlow250} & 1.46 & 7.12 & 6.15 & 0.00157\\
\textbf{MF-GP-nlow500} & 0.828 & 3.75 & 9.73 & 0.00167\\
\textbf{MF-GP-nlow1000} & 0.866 & 3.55 & 26.8 & 0.00236\\
\bottomrule
\end{tabular}
\end{table}

\par We then assessed the accuracy of the multi-fidelity GP models (section \ref{mf_gp}). All models used the 50 high-fidelity ($C_{T,LES}^*$) training points and a varying number of low-fidelity ($C_{T,wake}^*$) training points (using an ambient TI of 10\% for $C_{T,wake}^*$). The results from LOOCV are shown in table \ref{tab:nlow_sensitivity}. For the LOOCV we train our model on 49 out of the 50 high-fidelity data points and all low-fidelity data points. Then we average the error in predicting the high-fidelity data point left of the training set and repeat this in turn for data points. Increasing the number of low-fidelity training points from 250 to 500 reduced the mean and maximum error. However, increasing this to 1000 low-fidelity training points did not increase accuracy and increased the fitting and prediction time. This is because the number of high-fidelity training points is fixed. There is a threshold where the model of the relationship between $f_{LES}$ and $f_{wake}$, denoted $\rho$, limits the final accuracy of the emulator of $f_{LES}$. 

\begin{figure}
\centering
\includegraphics[width=0.8\textwidth]{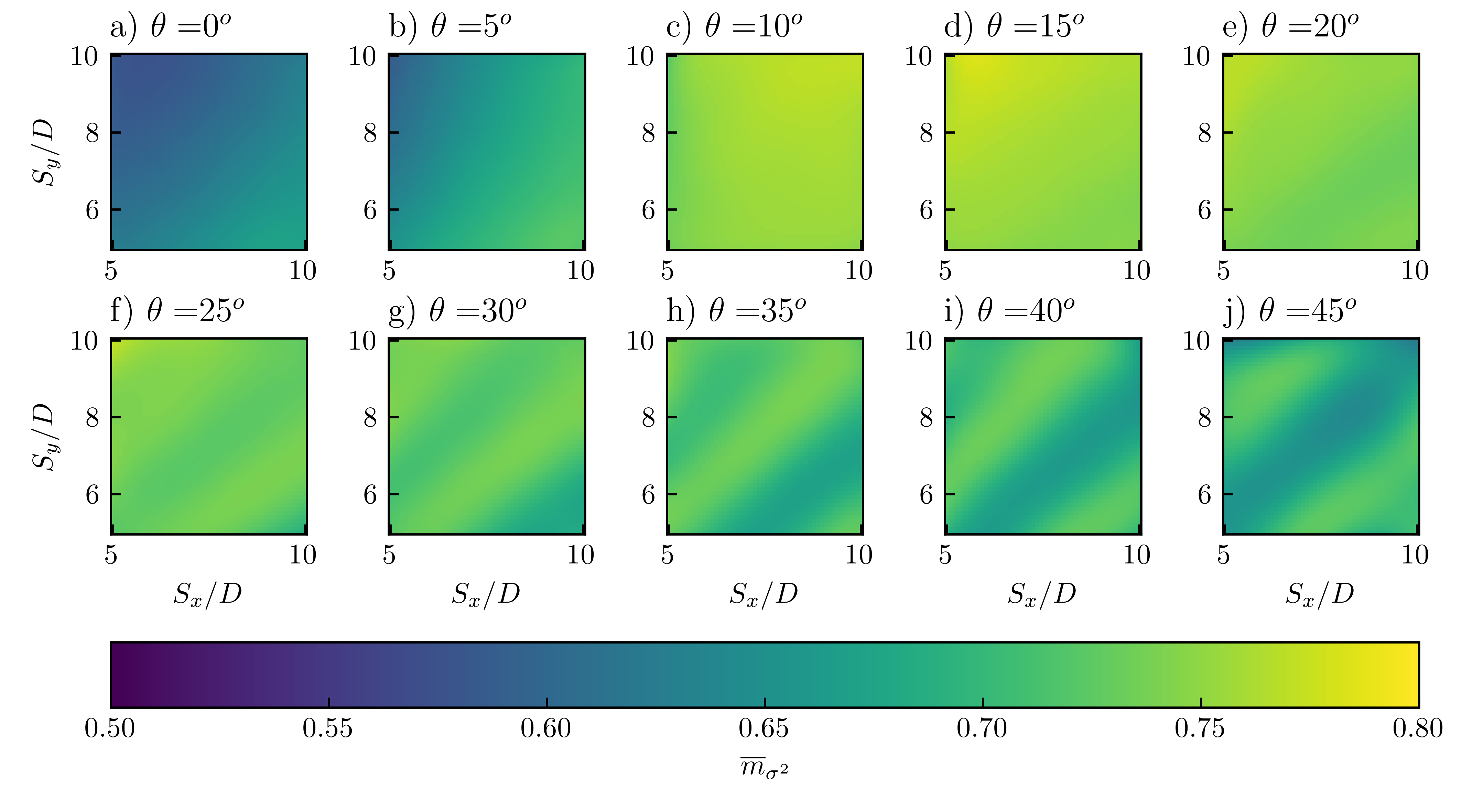}
\caption{Posterior mean function for $g_{high}(v)$ of \textbf{MF-GP-nlow500}.}
\label{fig:hf_posterior_mean}
\end{figure}

\par The posterior mean $\overline{m}_{\sigma^2}$ of $g_{low}(v)$ is an emulator of $f_{wake}$ and $g_{high}(v)$ is an emulator of $f_{LES}$. Figure \ref{fig:hf_posterior_mean} gives the predictions from the posterior mean of $g_{high}(v)$ (for \textbf{MF-GP-nlow500}). The lowest $\overline{m}_{\sigma^2}$ values were for a wind direction of $\theta=0^o$. $\overline{m}_{\sigma^2}$ increased rapidly with $\theta$ reaching a maximum of slightly over 0.75 at $\theta=10^o$. For large values of $\theta$ (above $\theta=25^o$) there were local minima in $\overline{m}_{\sigma^2}$ which appear in figure \ref{fig:hf_posterior_mean} as diagonal strips of low $\overline{m}_{\sigma^2}$ values. The main diagonal strip occurs along the line of $S_y=S_x\tan(\theta)$. There are two smaller strips either side of with positions given by $S_y=2\tan(\theta)$ and $S_y=0.5\tan(\theta)$ (this is discussed further in section \ref{section:discussion}).

\begin{figure}
\centering
\includegraphics[width=0.8\textwidth]{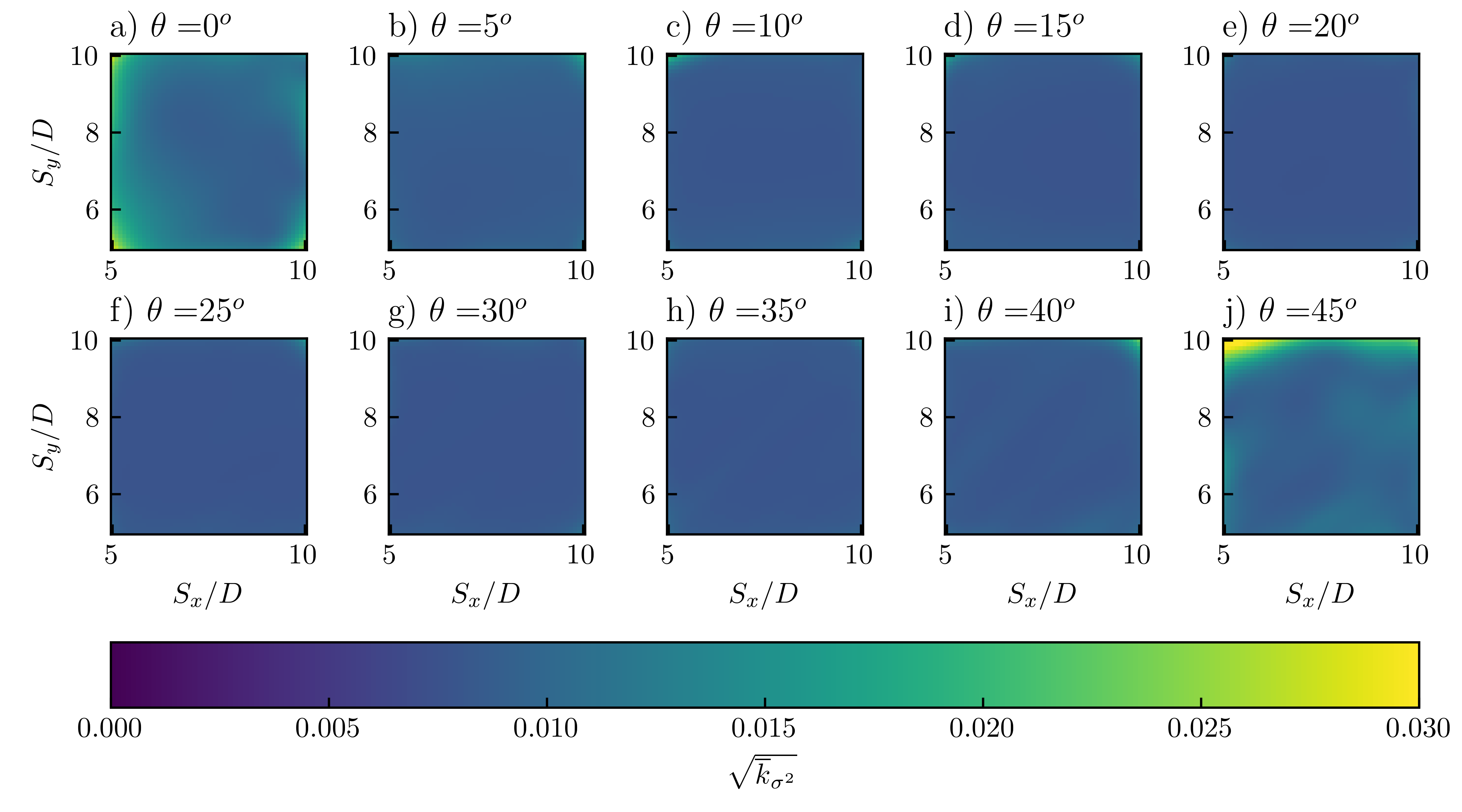}
\caption{Posterior variance function for $g_{high}(v)$ of \textbf{MF-GP-nlow500}.}
\label{fig:hf_posterior_variance}
\end{figure}

\par The uncertainty the model $\textbf{MF-GP-nlow500}$ has in predicting $f_{LES}$ is shown in figure \ref{fig:hf_posterior_variance}. The model uncertainty is uniform throughout the parameter space with slightly higher values at $\theta=0^o$ and $45^o$. Compared to the posterior variance of $\textbf{GP-wake-TI10-prior}$ (shown in figure \ref{fig:basic_gp_std}) the uncertainty is lower. By incorporating information from $C_{T,wake}^*$, the multi-fidelity GP model has more confidence about predicting $f_{LES}$.

\begin{figure}
\centering
\includegraphics[width=0.8\textwidth]{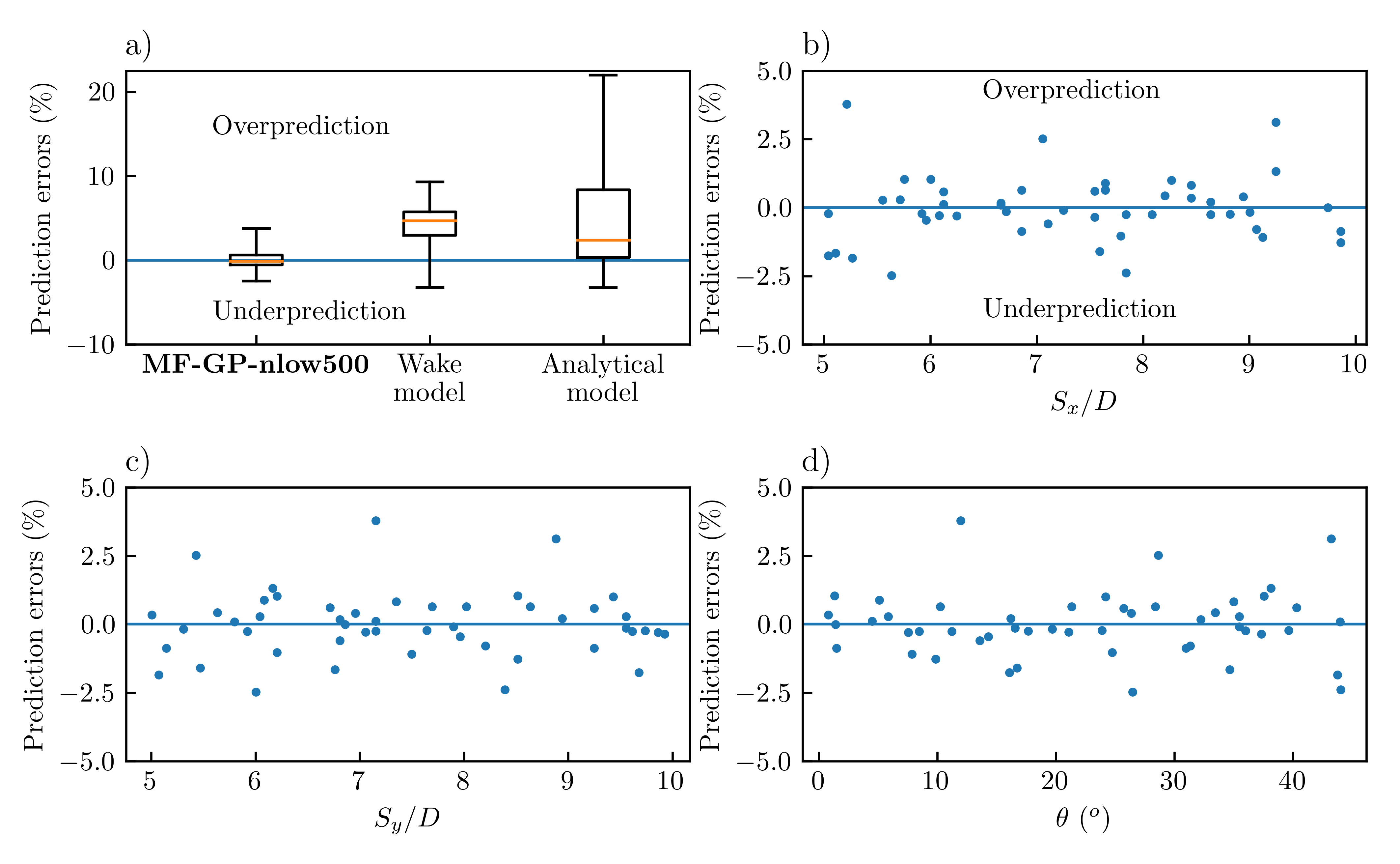}
\caption{Comparison of LOOCV prediction errors (\%) for different models a) and LOOCV prediction error (\%) of \textbf{MF-GP-nlow500} against input parameters b) $S_x/D$, c) $S_y/D$ and d) $\theta (^o)$. Note that for the box plot in a) the orange line is the median LOOCV error and the box is the interquartile range of LOOCV error.}
\label{fig:loocv_errors}
\end{figure}

\par The prediction errors from the LOOCV (for \textbf{MF-GP-nlow500}) are shown in figure \ref{fig:loocv_errors}. The box plot of prediction errors in figure \ref{fig:loocv_errors}a shows that this model had no significant bias whereas both the wake and analytical models systemically overestimated $C_{T,LES}^*$. Figures \ref{fig:loocv_errors}b-d show that for the statistical model there appears to be no part of the parameter space which had larger errors.

\begin{figure}
\centering
\includegraphics[width=0.8\textwidth]{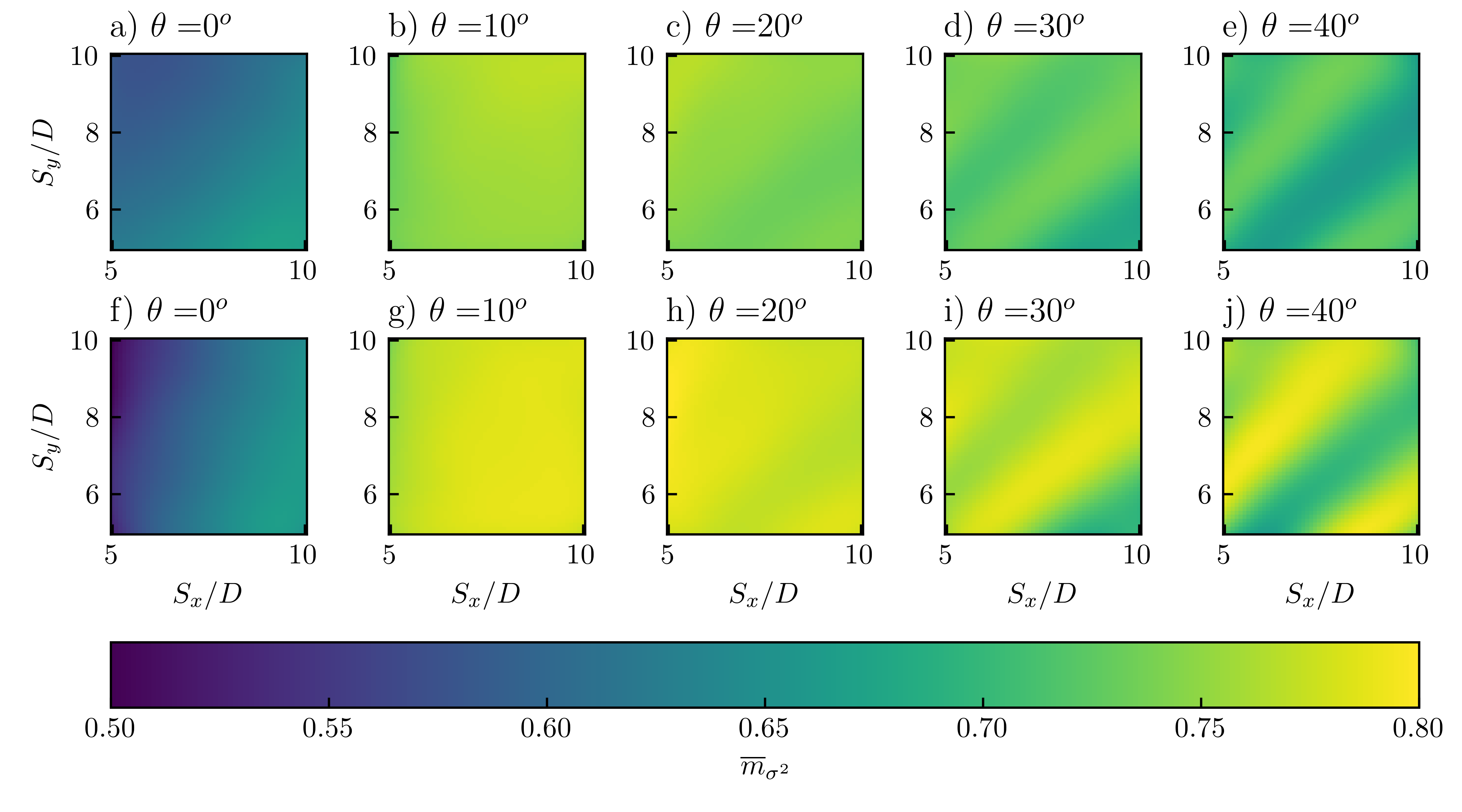}
\caption{Posterior mean function of \textbf{MF-GP-nlow500} for different values of $\theta$ for a) to e) $g_{high}(v)$ and f) to j) $g_{low}(v)$.}
\label{fig:mf_posterior_mean}
\end{figure}

The multi-fidelity approach used in this study builds a statistical model of both the low-fidelity ($f_{wake}$) and high-fidelity ($f_{LES}$) model. We can use the posterior means of $g_{low}(v)$ and $g_{high}(v)$ to see the differences between the wake model and LES. The posterior mean for both models are shown in figure \ref{fig:mf_posterior_mean}. For the wake model the change in $\overline{m}_{\sigma^2}$ with $\theta$ is greater than for the LES (especially between $\theta=0^o$ and $10^o$). For larger values of $\theta$, there is a larger difference in $\overline{m}_{\sigma^2}$ between waked and unwaked layouts for the low-fidelity model compared to the high-fidelity one. This suggests than the wake model is more sensitive to changes in wind directions than the LES.

\FloatBarrier

\subsection{Prediction of wind farm performance}\label{cp_results}

\par We use the predicted values of $C_{T,LES}^*$ from the emulators to predict the power output of wind farms under various mesoscale atmospheric conditions, following the concept of the two-scale momentum theory. We predict the (farm-averaged) turbine power coefficient $C_p$ using $C_{T,LES}^*$ predictions from \textbf{MF-GP-nlow500}. We call this prediction of farm performance $C_{p,model}$. Firstly, we use the $C_{T,LES}^*$ prediction from the LOOCV procedure as $C_T^*$ in equation \ref{windfarmmomentum} to calculate $\beta$ for a given value of wind extractability $\zeta$. We substitute this value of $\beta$ into the expression $C_p=\beta^3 {C_T^*}^\frac{3}{2}{C_T'}^{-\frac{1}{2}}$ (which is only valid for actuator discs) to calculate $C_{p,model}$. We compare the value of $C_{p,model}$ with the turbine power coefficient recorded in the LES, $C_{p,LES}$. The effect of the coarse LES resolution on turbine thrust (and hence also ABL response and $C_p$) has already been corrected \cite{Kirby2022}. The LES was performed with periodic horizontal boundary conditions and a fixed momentum supply, i.e., $\zeta=0$. However, the $C_{p,LES}$ has also been adjusted for a given $\zeta$ by scaling the velocity fields assuming Reynolds number independence\cite{Kirby2022}.

\par Similarly, the analytical model of $C_T^*$ can be used to give a theoretical prediction of wind farm performance called $C_{p,Nishino}$\cite{Kirby2022}, which is given by

\begin{equation}
    C_{p,Nishino} = \frac{64C_T'}{(4+C_T')^3} \left[ \frac{-\zeta + \sqrt{\zeta^2 + 4\left( \frac{16C_T'}{(4+C_T')^2}\frac{\lambda}{C_{f0}}+1\right)(1+\zeta)}}{2\left( \frac{16C_T'}{(4+C_T')^2}\frac{\lambda}{C_{f0}}+1\right)} \right]^3.
    \label{cp_nishino}
\end{equation}

\noindent We will compare the accuracy of both $C_{p,model}$ and $C_{p,Nishino}$ in predicting $C_{p,LES}$.

\par
\begin{figure}
\centering
\includegraphics[width=0.8\textwidth]{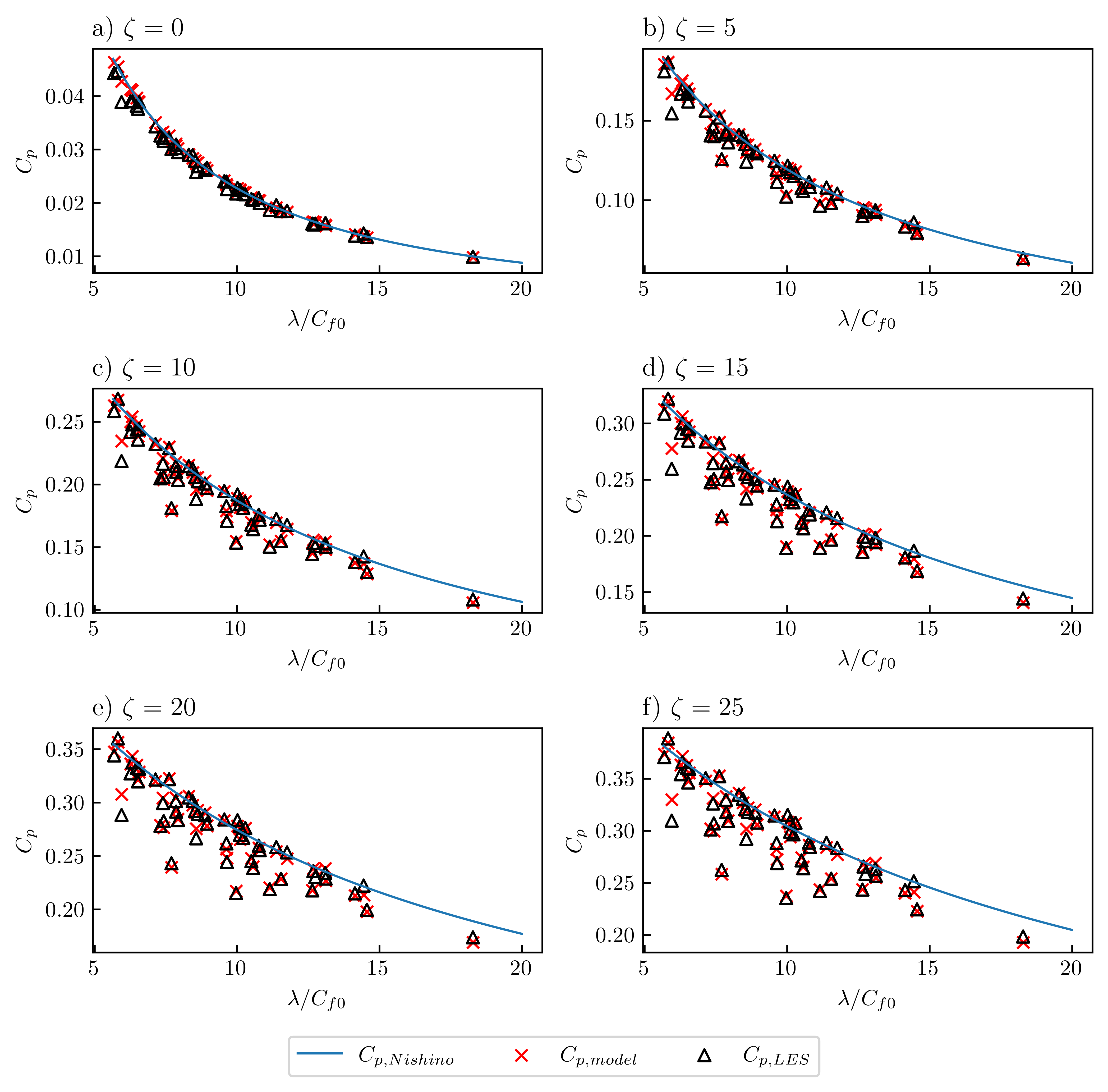}
\caption{Comparison of $C_p$ predictions with LES results for a realistic range of $\zeta$ values.}
\label{fig:cp_predictions}
\end{figure}

 Both $C_{p,model}$ and $C_{p,LES}$ are shown in figure \ref{fig:cp_predictions} for a realistic range of wind extractability factors, along with the results from $C_{p,Nishino}$ (equation \ref{cp_nishino}). $C_{p,Nishino}$ provides an approximate upper limit of farm-averaged $C_p$ as it predicts very well the effects of array density and large-scale atmospheric response. The statistical model accurately predicts the effect of turbine layout on farm performance which becomes more important with larger $\zeta$ values. As $\zeta$ increases, there is a larger difference between $C_{p,LES}$ and $C_{p,Nishino}$. Also, $C_{p,model}$ becomes slightly less accurate when $\zeta$ increases.

\par Table \ref{tab:cp_results} shows the average prediction errors of $C_{p,model}$ and $C_{p,Nishino}$. We quantified the mean absolute error using two different reference powers. Using $C_{p,LES}$ as the reference power, $C_{p,Nishino}$ had an error of around 5\% and the error increases with $\zeta$. The mean absolute error of $C_{p,model}$ was typically less than 1.5\% and this decreased slightly as $\zeta$ increases (due to the reference power $C_{p,LES}$ increasing). We also use the power of an isolated ideal turbine, $C_{p,Betz}$, as a reference power. $C_{p,Betz}$ is calculated using the actuator disc theory with the expression $C_{p,Betz}=64C_T'/(4+C_T')^3$ (note that in this study $C_T'=1.33$ and hence $C_{p,Betz}=0.563$). In this case the mean absolute error increased with $\zeta$ for both $C_{p,model}$ and $C_{p,Nishino}$. However, the average prediction error of $C_{p,model}$ remained below 0.65\%.

\begin{table}
\caption{Comparison of models for $C_p$ prediction.}
\label{tab:cp_results}
\centering
\begin{tabular}{ccc|ccc}
\toprule
\multicolumn{3}{c|}{$\frac{1}{50}\sum_{i=1}^{50}|C_{p,i}-C_{p,LES}|/C_{p,LES}$} & \multicolumn{3}{c}{$\frac{1}{50}\sum_{i=1}^{50}|C_{p,i}-C_{p,LES}|/C_{p,Betz}$}\\
\midrule
$\zeta$ & $C_{p,Nishino}$ & $C_{p,model}$ & $\zeta$ & $C_{p,Nishino}$ & $C_{p,model}$ \\
0 & 2.82\% & 2.15\% & 0 & 0.142\% & 0.108\%\\
5 & 4.38\% & 1.48\% & 5 & 0.954\% & 0.338\%\\
10 & 5.16\% & 1.35\% & 10 & 1.67\% & 0.459\%\\
15 & 5.66\% & 1.30\% & 15 & 2.24\% & 0.542\%\\
20 & 6.02\% & 1.26\% & 20 & 2.72\% & 0.601\%\\
25 & 6.30\% & 1.24\% & 25 & 3.11\% & 0.648\%\\
\bottomrule
\end{tabular}
\end{table}

\FloatBarrier

\section{Discussion}\label{section:discussion}

\par Data-driven modelling of the internal turbine thrust coefficient $C_T^*$ is a novel approach to modelling turbine-wake interactions. Data-driven models of wind farm performance typically focus on predicting the power output, which, however, depends on flow physics across a wide range of scales. Current data-driven approaches are either not generalisable to different atmospheric responses, or would require a very large set of expensive training data, such as finite-size wind farm LES data. Data-driven models of $C_T^*$ captures the effects of turbine-wake interactions, whilst also being applicable to different atmospheric responses (following the concept of the two-scale momentum theory).

\par The statistical emulator of $C_T^*$ developed in this study was able to predict the farm power $C_p$ of Kirby et. al.\cite{Kirby2022} with an average error of less than 0.65\%. The high accuracy and very low computational cost of this approach shows the potential of this approach for modelling turbine-wake interactions. It has several advantages over traditional approaches using the superposition of wake models. Information from turbulence-resolving LES is included which ensures a high accuracy. It will also be more advantageous as wind farms become larger because wake models struggle to capture the complex multi-scale flows physics which are important for large farms. The statistical model of $C_T^*$ may therefore allow fast and accurate predictions of wind farm performance.

\par All emulators developed in this study gave substantially better predictions of $C_{T,LES}^*$ compared to the analytical and wake models. Both the mean and maximum prediction errors were reduced by the emulators. The standard GP regression approach had a mean prediction error of 1.26\% and maximum error of approximately 6\%. The accuracy depends on the size of the LES data set and could be further decreased with a larger training set. The multi-fidelity GP approach gave more accurate predictions of $C_{T,LES}^*$ compared to the standard GP regression. This is because non-linear information fusion algorithm has incorporated information from many low-fidelity data points to improve the emulator of the high-fidelity (LES) model. This approach has the advantage that, unlike the standard GP regression approach, it is not necessary to evaluate the prior mean before making a prediction. Therefore, to predict $C_T^*$ it is only necessary to evaluate the posterior mean of the high-fidelity emulator for a specific turbine layout.

\par The shape of the posterior mean in figure \ref{fig:hf_posterior_mean} gives insights into the physics of turbine-wake interactions. This is because $C_{T,LES}^*$ is low when a layout has a high degree of turbine-wake interactions. For the turbine operating conditions used, $C_{T,LES}^*$ is close to 0.75 when a layout has a small degree of wake interactions. Figure \ref{fig:hf_posterior_mean}a shows $C_{T,LES}^*$ when the wind direction is perfectly aligned with the rows of turbines ($\theta=0$). This gives wind farms with a high degree of wake interactions which results in low $C_{T,LES}^*$ values. For $\theta=0^o$, increasing $S_x/D$ increases $C_T^*$ because there is a larger streamwise distance between turbines for the wakes to recover. When the cross-streamwise spacing ($S_y/D$) is increased the degree of wake interactions increases, i.e., $C_{T,LES}^*$ decreases. This is because there is a lower array density which results in a lower turbulence intensity within the farm and hence slower wake recovery. Yang\cite{Yang2012} found that increasing the cross-streamwise spacing in infinitely-large wind farms increased the power of individual turbines and concluded that this was due to reduced wake interactions. However, the increase in turbine power found by Yang\cite{Yang2012} may be also explained by to a faster farm-averaged wind speed caused by a reduced array density rather than reduced wake interactions.

\par
\begin{figure}
\centering
\includegraphics[width=0.8\textwidth]{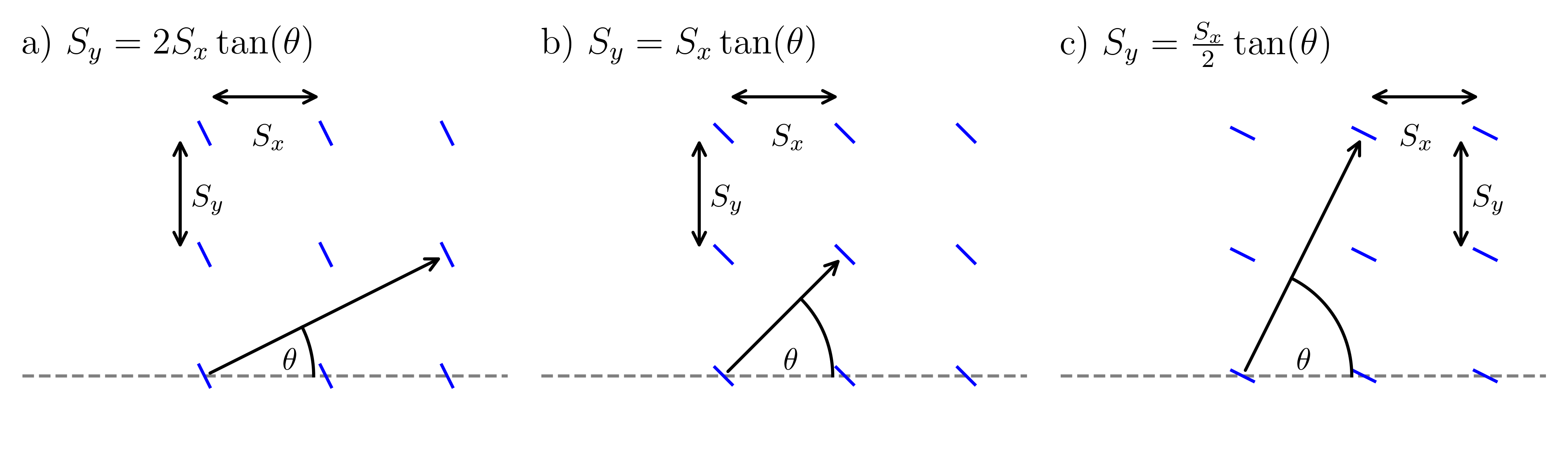}
\caption{Alignment of turbines for different combinations of $S_x$, $S_y$ and $\theta$.}
\label{fig:angles}
\end{figure}

 When the wind direction $\theta$ increases, $C_{T,LES}^*$ increases to a maximum of just over 0.75 at $\theta=10^o$ (figure \ref{fig:hf_posterior_mean}c). This result agrees qualitatively with another study\cite{Stevens2014} in which it was found that the maximum farm power was produced by an intermediate wind direction. When $\theta$ increases above $20^o$ regions of low $C_{T,LES}^*$ appear diagonally (see figures \ref{fig:hf_posterior_mean}f-j). The regions of low $C_{T,LES}^*$ are centred on the surfaces given by $S_y=2S_x\tan(\theta)$, $S_y=S_x\tan(\theta)$ and $S_y=0.5S_x\tan(\theta)$. These regions correspond to turbines being aligned along different axes throughout the farm (see figure \ref{fig:angles}). There are longer streamwise distance between turbines for these arrangements (compared to $\theta=0^o$) and so the $C_{T,LES}^*$ values are higher than for $\theta=0^o$.

\par The accuracy of the statistical emulators could be further improved in future studies. Both the standard and multi-fidelity GP models can be improved by adding more evaluations of $C_{T,LES}^*$. From table \ref{tab:nlow_sensitivity}, the accuracy of the multi-fidelity GP models did not improve once we used more than 500 $C_{T,wake}^*$ evaluations. This shows that the error in predicting $C_{T,LES}^*$ for \textbf{MF-GP-nlow500} is not due to the model of $f_{wake}$. Instead the error arises from the learnt relationship between $f_{wake}$ and $f_{LES}$.

\par The statistical emulators developed are not applicable to all wind farms because of the limited nature of our data set. A limitation of the developed model is that it is only applicable to farms with perfectly aligned layouts. It should also be noted that our model was trained on data from simulations of a neutrally stratified boundary layer. Therefore a larger LES data set with an extended parameter space would be required to account for the effect of atmospheric stability on wake interactions and the resulting $C_T^*$. Another limitation of our model is that it assumes all turbines have the same resistance coefficient $C_T'$. It is likely that this condition can be strictly satisfied only in the fully developed region of a large farm where the wind speed does not change in the streamwise or cross-streamwise directions.

\par Although we considered only actuator discs in this study for demonstration, the proposed approach using a data-driven model of $C_T*$ can be applied to power prediction of real turbines as well in future studies. In this study, we calculate $C_{p,model}$ using the expression $C_{p,model}=\beta^3 {C_T^*}^\frac{3}{2}{C_T'}^{-\frac{1}{2}}$. This assumes that the relationship between $C_p^*$ and $C_T'$ is given by $C_p^*={C_T^*}^\frac{3}{2}{C_T'}^{-\frac{1}{2}}$, which is only valid for actuator discs. For real turbines, the relationship between $C_p^*$ and $C_T'$ can be calculated using BEM theory\cite{Nishino2018} according to the turbine design and operating conditions (noting that the turbine induction factor can still be estimated as $a=C_T'/(4+C_T')$). $C_{p,model}$ can then be calculated using equation \ref{cp} with $\beta$ found using equation \ref{windfarmmomentum}. However, for a data-driven model of $C_T^*$ to be applicable to real turbines, it will be necessary to model the impact of a variable $C_T'$ rather than assuming a fixed $C_T'$ value as in this study.
 
\section{Conclusions}\label{section:conclusion}

\par In this study we proposed a new data-driven approach to modelling turbine wake interactions and resulting flow resistance in large wind farms. We developed statistical emulators of the farm-internal turbine thrust coefficient $C_{T,LES}^*$ as a function of turbine layout and wind direction. $C_T^*$ represents the flow resistance within a wind farm and reflects the characteristics of the turbine-scale flows including wake and turbine blockage effects. We developed several emulators using both standard GP regression and multi-fidelity GP regression. The standard GP was trained using data from 50 infinitely-large wind farm LES (and using a low-fidelity wake model as a prior mean). The multi-fidelity GP was trained using data from both LES and wake model simulations. We estimated the test accuracy of the model by performing leave-one-out cross-validation and assessed the error in predicting $C_{T,LES}^*$. All emulators had a mean test error of less than 2\% for predicting $C_{T,LES}^*$. The multi-fidelity GP gave the best performance with a mean prediction error of 0.849\% and maximum prediction error of 3.78\% with no bias for under or over-prediction. This is low compared to the mean error of the wake model (4.60\%) and analytical $C_T^*$ model (5.26\%) which both had a bias for overpredicting $C_{T,LES}^*$. 

\par We used an emulator of $C_{T,LES}^*$ to make predictions of wind farm performance under various mesoscale atmospheric conditions (characterised by the wind extractability factor $\zeta$) using the two-scale momentum theory \cite{Nishino2020}. Our predictions of farm power production had an average error of less than 1.5\% under realistic wind extractability scenarios compared to the LES. When the error in power prediction is expressed relative to the power of an isolated ideal turbine the average prediction error is less than 0.7\%. We also used a previously proposed analytical model of $C_T^*$ \cite{Nishino2016} to predict farm power output with an average error of less than 3.5\% (with the power of an isolated turbine as the reference power). The analytical model correctly predicts the trends in farm performance with array density under different scenarios of large-scale atmospheric response, although it tends to overpredict the power where turbine-wake interactions are important. Using statistical emulators of $C_T^*$ is a new approach to modelling turbine-wake interactions and flow resistance within large wind farms. The approach can be extended in future studies by increasing the size of the training data set, for example, to account for the effects of $C_T'$ and atmospheric stability conditions on $C_T^*$. The very low computational cost and high accuracy of the model could be beneficial for future wind farm optimisation.

\section*{Acknowledgments}
The first author (AK) acknowledges the NERC-Oxford Doctoral Training Partnership in
Environmental Research (NE/S007474/1) for funding and training. 

\subsection*{Author contributions}
T.N. derived the theory. A.K. and T.D.D. performed the simulations. F-X.B. provided assistance and guidance for the machine learning methodology. A.K. wrote the paper with corrections from T.N., F-X.B and T.D.D.

\subsection*{Financial disclosure}
None reported.

\subsection*{Conflict of interest}
The authors report no conflict of interest.

\subsection*{Data availability statement}
The data and code that support the findings of this study are openly available at
\url{https://github.com/AndrewKirby2/ctstar_statistical_model}. This includes the results from the wind farm LES and wake model simulations. The repository also includes the code for the results presented in sections \ref{standardgp}, \ref{results_ct*} and \ref{cp_results}.

\subsection*{Author ORCID}
A. Kirby, \url{https://orcid.org/0000-0001-8389-1619}; F-X. Briol \url{https://orcid.org/0000-0002-0181-2559}; T. Nishino, \url{https://orcid.org/0000-0001-6306-7702}.

\bibliography{references}

\begin{thebibliography}{10}
\providecommand \doibase [0]{http://dx.doi.org/}%

\bibitem{Porte-Agel2020}
Porté-Agel F, Bastankhah M, Shamsoddin S. Wind-Turbine and Wind-Farm Flows: A
  Review. {\it Boundary-Layer Meteorology} 2020\string; 174\string: 1-59.
\newblock \href {\doibase 10.1007/s10546-019-00473-0} {doi:
  10.1007/s10546-019-00473-0}

\bibitem{Bleeg2018}
Bleeg J, Purcell M, Ruisi R, Traiger E. Wind farm blockage and the consequences
  of neglecting its impact on energy production. {\it Energies} 2018\string;
  11\string: 1609.
\newblock \href {\doibase 10.3390/en11061609} {doi: 10.3390/en11061609}

\bibitem{CarbonTrust2022}
{Carbon Trust} . Global Blockage Effect in Offshore Wind (GloBE) [accessed
  07/11/2022].
  \url{https://www.carbontrust.com/our-projects/large-scale-rd-projects-offshore-wind/global-blockage-effect-in-offshore-wind-globe};
  2022.

\bibitem{Jensen1983}
Jensen NO. A note on wind generator interaction. {\it Risø-M-2411 Risø
  National Laboratory Roskilde} 1983.

\bibitem{Bastankhah2014}
Bastankhah M, Porté-Agel F. A new analytical model for wind-turbine wakes.
  {\it Renewable Energy} 2014\string; 70\string: 116-123.
\newblock \href {\doibase 10.1016/j.renene.2014.01.002} {doi:
  10.1016/j.renene.2014.01.002}

\bibitem{Katic1986}
Katic I, Hojstrup J, Jensen NO. A simple model for cluster efficiency. {\it
  Proceedings of the European wind energy association conference and
  exhibition, Rome, Italy} 1986\string: 407-409.

\bibitem{Zong2020}
Zong H, Porté-Agel F. A momentum-conserving wake superposition method for wind
  farm power prediction. {\it Journal of Fluid Mechanics} 2020\string;
  889\string: A8.
\newblock \href {\doibase 10.1017/jfm.2020.77} {doi: 10.1017/jfm.2020.77}

\bibitem{Kirby2022}
Kirby A, Nishino T, Dunstan TD. Two-scale interaction of wake and blockage
  effects in large wind farms. {\it Journal of Fluid Mechanics} 2022\string;
  953\string: A39.
\newblock \href {\doibase 10.1017/jfm.2022.979} {doi: 10.1017/jfm.2022.979}

\bibitem{Stevens2016a}
Stevens RJAM, Gayme DF, Meneveau C. Effects of turbine spacing on the power
  output of extended wind-farms. {\it Wind Energy} 2016\string; 19\string:
  359-370.
\newblock \href {\doibase 10.1002/we.1835} {doi: 10.1002/we.1835}

\bibitem{Fitch2012}
Fitch AC, Olson JB, Lundquist JK, et al. Local and mesoscale impacts of wind
  farms as parameterized in a mesoscale NWP model. {\it Monthly Weather Review}
  2012\string; 140.
\newblock \href {\doibase 10.1175/MWR-D-11-00352.1} {doi:
  10.1175/MWR-D-11-00352.1}

\bibitem{Abkar2015}
Abkar M, Porté-Agel F. A new wind-farm parameterization for large-scale
  atmospheric models. {\it Journal of Renewable and Sustainable Energy}
  2015\string; 7.
\newblock \href {\doibase 10.1063/1.4907600} {doi: 10.1063/1.4907600}

\bibitem{Pan2018}
Pan Y, Archer CL. A Hybrid Wind-Farm Parametrization for Mesoscale and Climate
  Models. {\it Boundary-Layer Meteorology} 2018\string; 168\string: 469-495.
\newblock \href {\doibase 10.1007/s10546-018-0351-9} {doi:
  10.1007/s10546-018-0351-9}

\bibitem{Zehtabiyan-Rezaie2022}
Zehtabiyan-Rezaie N, Iosifidis A, Abkar M. Data-driven fluid mechanics of wind
  farms: A review. {\it Journal of Renewable and Sustainable Energy}
  2022\string; 14\string: 32703.
\newblock \href {\doibase 10.1063/5.0091980} {doi: 10.1063/5.0091980}

\bibitem{Renganathan2022}
Renganathan SA, Maulik R, Letizia S, Iungo GV. Data-driven wind turbine wake
  modeling via probabilistic machine learning. {\it Neural Computing and
  Applications} 2022\string; 34\string: 6171-6186.
\newblock \href {\doibase 10.1007/s00521-021-06799-6} {doi:
  10.1007/s00521-021-06799-6}

\bibitem{Optis2019}
Optis M, Perr-Sauer J. The importance of atmospheric turbulence and stability
  in machine-learning models of wind farm power production. {\it Renewable and
  Sustainable Energy Reviews} 2019\string; 112\string: 27-41.
\newblock \href {\doibase 10.1016/j.rser.2019.05.031} {doi:
  10.1016/j.rser.2019.05.031}

\bibitem{Japar2014}
Japar F, Mathew S, Narayanaswamy B, Lim CM, Hazra J. Estimating the wake losses
  in large wind farms: A machine learning approach. {\it ISGT 2014}
  2014\string: 1-5.
\newblock \href {\doibase 10.1109/ISGT.2014.6816427} {doi:
  10.1109/ISGT.2014.6816427}

\bibitem{Yan2019}
Yan C, Pan Y, Archer CL. A general method to estimate wind farm power using
  artificial neural networks. {\it Wind Energy} 2019\string; 22\string:
  1421-1432.
\newblock \href {\doibase 10.1002/we.2379} {doi: 10.1002/we.2379}

\bibitem{Zhang2022}
Zhang J, Zhao X. Wind farm wake modeling based on deep convolutional
  conditional generative adversarial network. {\it Energy} 2022\string;
  238\string: 121747.
\newblock \href {\doibase https://doi.org/10.1016/j.energy.2021.121747} {doi:
  https://doi.org/10.1016/j.energy.2021.121747}

\bibitem{Wilson2017}
Wilson B, Wakes S, Mayo M. Surrogate modeling a computational fluid
  dynamics-based wind turbine wake simulation using machine learning. {\it 2017
  IEEE Symposium Series on Computational Intelligence (SSCI)} 2017\string: 1-8.
\newblock \href {\doibase 10.1109/SSCI.2017.8280844} {doi:
  10.1109/SSCI.2017.8280844}

\bibitem{Ti2020}
Ti Z, Deng XW, Yang H. Wake modeling of wind turbines using machine learning.
  {\it Applied Energy} 2020\string; 257\string: 114025.
\newblock \href {\doibase https://doi.org/10.1016/j.apenergy.2019.114025} {doi:
  https://doi.org/10.1016/j.apenergy.2019.114025}

\bibitem{Ti2021}
Ti Z, Deng XW, Zhang M. Artificial Neural Networks based wake model for power
  prediction of wind farm. {\it Renewable energy} 2021\string; 172\string:
  618-631.
\newblock \href {\doibase https://doi.org/10.1016/j.renene.2021.03.030} {doi:
  https://doi.org/10.1016/j.renene.2021.03.030}

\bibitem{Park2019}
Park J, Park J. Physics-induced graph neural network: An application to
  wind-farm power estimation. {\it Energy} 2019\string; 187.
\newblock \href {\doibase 10.1016/j.energy.2019.115883} {doi:
  10.1016/j.energy.2019.115883}

\bibitem{Bleeg2020}
Bleeg J. A Graph Neural Network Surrogate Model for the Prediction of Turbine
  Interaction Loss. {\it Journal of Physics: Conference Series} 2020\string;
  1618.
\newblock \href {\doibase 10.1088/1742-6596/1618/6/062054} {doi:
  10.1088/1742-6596/1618/6/062054}

\bibitem{Nishino2020}
Nishino T, Dunstan TD. Two-scale momentum theory for time-dependent modelling
  of large wind farms. {\it Journal of Fluid Mechanics} 2020\string;
  894\string: A2.
\newblock \href {\doibase 10.1017/jfm.2020.252} {doi: 10.1017/jfm.2020.252}

\bibitem{Nishino2016}
Nishino T. Two-scale momentum theory for very large wind farms. {\it Journal of
  Physics: Conference Series} 2016\string; 753\string: 032054.
\newblock \href {\doibase 10.1088/1742-6596/753/3/032054} {doi:
  10.1088/1742-6596/753/3/032054}

\bibitem{Patel2021}
Patel K, Dunstan TD, Nishino T. Time-dependent upper limits to the performance
  of large wind farms due to mesoscale atmospheric response. {\it Energies}
  2021\string; 14\string: 6437.
\newblock \href {\doibase 10.3390/en14196437} {doi: 10.3390/en14196437}

\bibitem{Sacks1989}
Sacks J, Welch WJ, Mitchell TJ, Wynn HP. Design and analysis of computer
  experiments. {\it Statistical Science} 1989\string; 4\string: 409-423.
\newblock \href {\doibase 10.1214/ss/1177012413} {doi: 10.1214/ss/1177012413}

\bibitem{Currin1991}
Currin C, Mitchell T, Morris M, Ylvisaker D. Bayesian prediction of
  deterministic functions, with applications to the design and analysis of
  computer experiments. {\it Journal of the American Statistical Association}
  1991\string; 86\string: 953-963.
\newblock \href {\doibase 10.1080/01621459.1991.10475138} {doi:
  10.1080/01621459.1991.10475138}

\bibitem{Johnson1990}
Johnson ME, Moore LM, Ylvisaker D. Minimax and maximin distance designs. {\it
  Journal of Statistical Planning and Inference} 1990\string; 26\string:
  131-148.
\newblock \href {\doibase 10.1016/0378-3758(90)90122-B} {doi:
  10.1016/0378-3758(90)90122-B}

\bibitem{Santner2018}
Santner TJ, Williams BJ, Notz W. {\it The design and analysis of computer
  experiments}.
\newblock second~ed. 2018.

\bibitem{Wynne2021}
Wynne G, Briol FX, Girolami M. Convergence guarantees for gaussian process
  means with misspecified likelihoods and smoothness. {\it Journal of Machine
  Learning Research} 2021\string; 22.

\bibitem{Shapiro2019}
Shapiro CR, Gayme DF, Meneveau C. Filtered actuator disks: Theory and
  application to wind turbine models in large eddy simulation. {\it Wind
  Energy} 2019\string; 22\string: 1414-1420.
\newblock \href {\doibase 10.1002/we.2376} {doi: 10.1002/we.2376}

\bibitem{Niayifar2016}
Niayifar A, Porté-Agel F. Analytical modeling of wind farms: A new approach
  for power prediction. {\it Energies} 2016\string; 9.
\newblock \href {\doibase 10.3390/en9090741} {doi: 10.3390/en9090741}

\bibitem{pywake2.2.0_2020}
Pedersen MM, Laan v.~dP, Friis-Møller M, Rinker J, Réthoré PE.
  DTUWindEnergy/PyWake: PyWake.  2021.
\newblock \href {\doibase 10.5281/zenodo.2562662} {doi: 10.5281/zenodo.2562662}

\bibitem{Crespo1996}
Crespo A, Hernández J. Turbulence characteristics in wind-turbine wakes. {\it
  Journal of Wind Engineering and Industrial Aerodynamics} 1996\string;
  61\string: 71-85.
\newblock \href {\doibase 10.1016/0167-6105(95)00033-X} {doi:
  10.1016/0167-6105(95)00033-X}

\bibitem{Rasmussen2018}
Rasmussen CE, Williams CKI. {\it Gaussian Processes for Machine Learning}.
\newblock the MIT Press .
\newblock 2018

\bibitem{Peherstorfer2018}
Peherstorfer B, Willcox K, Gunzburger M. Survey of multifidelity methods in
  uncertainty propagation, inference, and optimization. {\it SIAM Review}
  2018\string; 60.
\newblock \href {\doibase 10.1137/16M1082469} {doi: 10.1137/16M1082469}

\bibitem{Perdikaris2017}
Perdikaris P, Raissi M, Damianou A, Lawrence ND, Karniadakis GE. Nonlinear
  information fusion algorithms for data-efficient multi-fidelity modelling.
  {\it Proceedings of the Royal Society A: Mathematical, Physical and
  Engineering Sciences} 2017\string; 473.
\newblock \href {\doibase 10.1098/rspa.2016.0751} {doi: 10.1098/rspa.2016.0751}

\bibitem{emukit2019}
Paleyes A, Pullin M, Mahsereci M, Lawrence N, González J. Emulation of
  physical processes with Emukit.  2019.

\bibitem{gpy2014}
{GPy} . {GPy}: A Gaussian process framework in python.
  \url{http://github.com/SheffieldML/GPy};  since 2012.

\bibitem{Yang2012}
Yang X, Kang S, Sotiropoulos F. Computational study and modeling of turbine
  spacing effects infinite aligned wind farms. {\it Physics of Fluids}
  2012\string; 24\string: 11510.
\newblock \href {\doibase 10.1063/1.4767727} {doi: 10.1063/1.4767727}

\bibitem{Stevens2014}
Stevens RJAM, Gayme DF, Meneveau C. Large eddy simulation studies of the
  effects of alignment and wind farm length. {\it Journal of Renewable and
  Sustainable Energy} 2014\string; 6\string: 023105.
\newblock \href {\doibase 10.1063/1.4869568} {doi: 10.1063/1.4869568}

\bibitem{Nishino2018}
Nishino T, Hunter W. {Tuning turbine rotor design for very large wind farms}.
  {\it Proceedings of the Royal Society A: Mathematical, Physical and
  Engineering Sciences} 2018\string; 474(2220)\string: 1--20.
\newblock \href {\doibase 10.1098/rspa.2018.0237} {doi: 10.1098/rspa.2018.0237}

\end{thebibliography}

\end{document}